%% file: main.tex
\documentclass[lettersize,journal]{IEEEtran}
\usepackage{diagbox}
\usepackage{amsthm}
\usepackage{amsmath,amsfonts}
\usepackage{algorithmic}
\usepackage{algorithm}
\usepackage{array}
\usepackage{textcomp}
\usepackage{stfloats}
\usepackage{hyperref}   
\usepackage{xurl}       
\usepackage{verbatim}
\usepackage{graphicx}
\usepackage{cite}
\usepackage{flushend}
\usepackage[caption=false,font=footnotesize]{subfig}
\usepackage{float}
\usepackage{subfig}
\usepackage{textcomp}
\usepackage{xcolor} 
\usepackage[printonlyused]{acronym}
\usepackage{multirow}
\newtheorem{theorem}{Theorem}

\providecommand{\lemmaname}{Lemma}
\providecommand{\theoremname}{Theorem}
\usepackage{color}
\usepackage{soul}
\usepackage{balance}
\renewcommand\color[2]{#2}

\usepackage{enumitem}

\begin{document}
\providecommand{\lemmaname}{Lemma}
\providecommand{\theoremname}{Theorem}
\newacro{P2P}{Peer to Peer}
\newacro{PBFT}{Practical Byzantine Fault Tolerance}
\newacro{FBFT}{Fast Byzantine Fault Tolerance}
\newacro{LCE}{Linked Cross-shard Endorsement} 
\newacro{SGX}{Software Guard Extensions}
\newacro{PoW}{Proof of Work}
\newacro{IoT}{Internet of Things}
\newacro{TEE}{Trusted Execution Environment}
\newacro{AWS}{Amazon Web Services} 
\newacro{TC}{Traffic Control} 
\newacro{BLS}{Boneh-Lynn-Shacham} 

\title{SpiralShard: Highly Concurrent and Secure Blockchain Sharding via Linked Cross-shard Endorsement}

\author{You~Lin,
        Mingzhe~Li \IEEEmembership{Member,~IEEE},
        and~Jin~Zhang,~\IEEEmembership{Member,~IEEE}
        
        \IEEEcompsocitemizethanks{
        \IEEEcompsocthanksitem This work was supported by the National Natural Science Foundation of China (NSFC) under Grant T2495254 and Fang Keng Fellowship.
        \IEEEcompsocthanksitem Y. Lin is with the Department of Computer Science and Engineering, Southern University of Science and Technology, Shenzhen 518055, China (email: liny2021@mail.sustech.edu.cn).
        \IEEEcompsocthanksitem M. Li is  with the Institute of High Performance Computing, A*STAR, Singapore, and with the School of Computing and Information Technology, Great Bay University, Dongguan 523000, China (email: mlibn@connect.ust.hk).
        \IEEEcompsocthanksitem J. Zhang is with the Department of Computer Science and Engineering
and the Research Institute of Trustworthy Autonomous Systems, Southern
University of Science and Technology, Shenzhen 518055, China, and also
with Peng Cheng Laboratory, Shenzhen 518055, China (email: zhangj4@sustech.edu.cn).
        \IEEEcompsocthanksitem Y. Lin and M. Li are the co-first authors.
        \IEEEcompsocthanksitem J. Zhang is the corresponding author.
        }
}

\markboth{IEEE/ACM Transactions on Networking, VOL. XX, NO. XX, XX 2024}%
{Lin \MakeLowercase{\textit{et al.}}: SpiralShard: Highly Concurrent and Secure Blockchain Sharding via Linked Cross-shard Endorsement}


\maketitle

\input{Files/abstract}


\input{Files/introduction}

\input{Files/relatedWork_motivation}

\input{Files/model}

\input{Files/system_overview}

\input{Files/protocol_desigN}

\input{Files/analysis}

\input{Files/evaluation}
\input{Files/conclusion}

\bibliographystyle{abbrv}
\bibliography{ref}

\vspace{-30pt}
\begin{IEEEbiography}[{\includegraphics[width=1in,height=1.25in,clip,keepaspectratio]{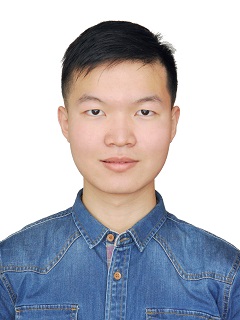}}]{You Lin}
is currently a master candidate with Department of Computer Science and Engineering, Southern University of Science and Technology. 
He received his B.E. degree in computer science and technology from Southern University of Science and Technology in 2021. 
His research interests are mainly in blockchain, network economics, and consensus protocols.
\end{IEEEbiography}
\vspace{-30pt}
\begin{IEEEbiography}[{\includegraphics[width=1in,height=1.25in,clip,keepaspectratio]{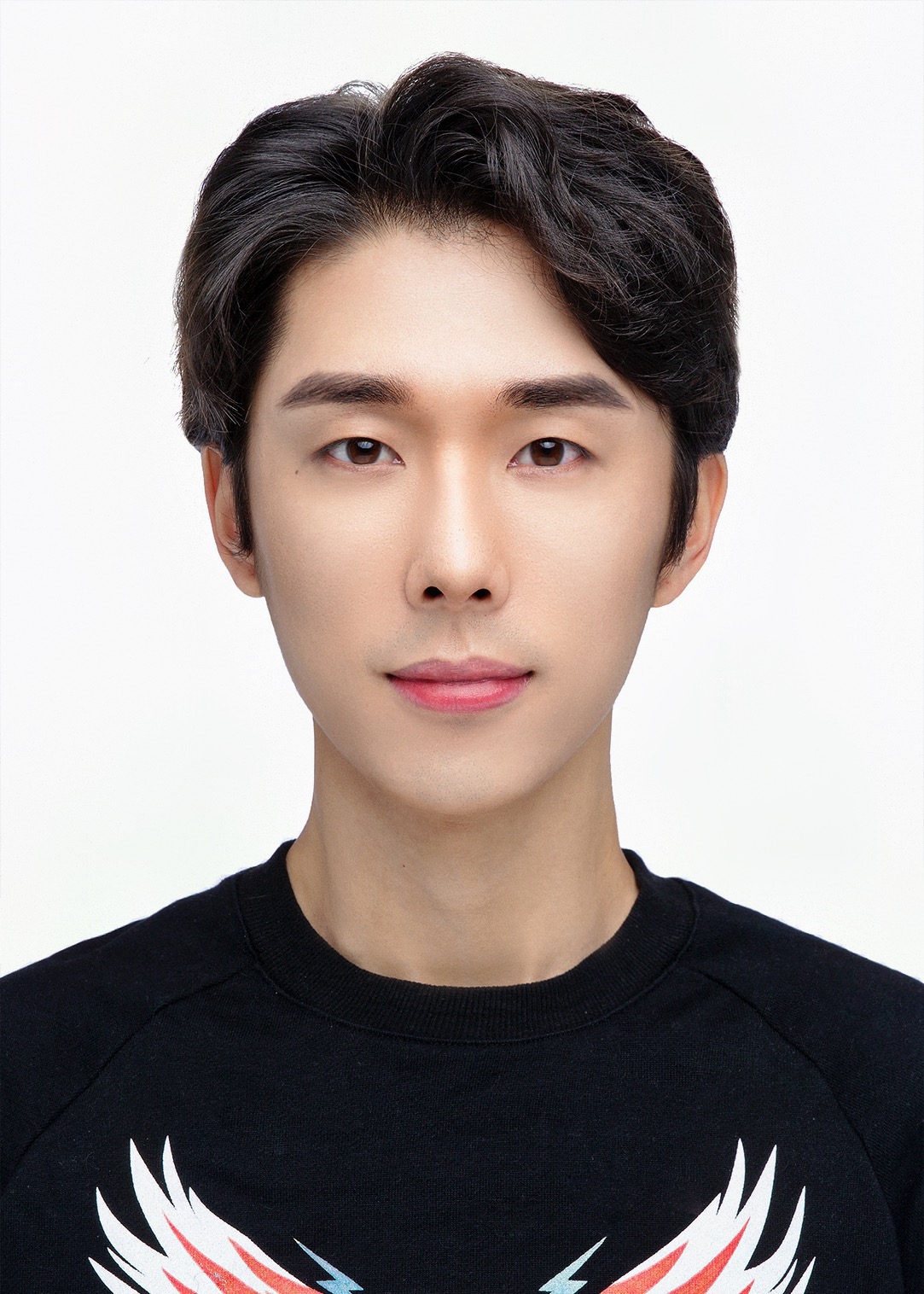}}]{Mingzhe Li}
is currently a Scientist with the Institute of High Performance Computing (IHPC), A*STAR, Singapore.
He received his Ph.D. degree from the Department of Computer Science and Engineering, Hong Kong University of Science and Technology in 2022.
Prior to that, he received his B.E. degree from Southern University of Science and Technology.
His research interests are mainly in blockchain sharding, consensus protocol, blockchain application, network economics, and crowdsourcing.
\end{IEEEbiography}
\vspace{-30pt}
\begin{IEEEbiography}
[{\includegraphics[width=1in,height=1.25in,clip,keepaspectratio]{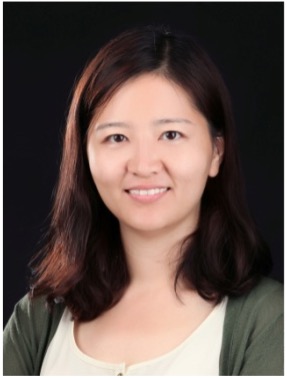}}]{Jin Zhang} 
is currently an associate professor with Department of Computer Science and Engineering, Southern University of Science and Technology. 
She received her B.E. and M.E. degrees in electronic engineering from Tsinghua University in 2004 and 2006, respectively, and received her Ph.D. degree in computer science from Hong Kong University of Science and Technology in 2009. 
Her research interests are mainly in mobile healthcare and wearable computing, wireless communication and networks, network economics, cognitive radio networks and dynamic spectrum management. 
\end{IEEEbiography}


 




\vfill

\end{document}

%% file: Files/abstract.tex
\begin{abstract}
Blockchain sharding improves the scalability of blockchain systems by partitioning the whole blockchain state, nodes, and transaction workloads into different shards. 
However, existing blockchain sharding systems generally suffer from a small number of shards, resulting in \emph{limited concurrency}. 
The main reason is that existing sharding systems require \emph{large shard sizes} to ensure security.

To enhance the concurrency of blockchain sharding securely, we propose SpiralShard. 
The intuition is to allow the existence of some shards with a larger fraction of malicious nodes (i.e., corrupted shards), 
thus reducing shard sizes. 
SpiralShard can configure more and smaller shards for higher concurrency at the same network size.
To ensure security with the existence of corrupted shards, we propose the Linked Cross-shard Endorsement (LCE) protocol. 
According to our LCE protocol, the blocks of each shard are sequentially verified and endorsed (via intra-shard consensus) by a group of shards before being finalized.
As a result, a corrupted shard can eliminate forks with the help of the other shards.
We implement SpiralShard based on Harmony and conduct extensive evaluations. 
Experimental results show that, compared with Harmony, SpiralShard achieves around 19$\times$ throughput gain under a large network size with 4,000+ nodes.
\end{abstract}

%% file: Files/introduction.tex
\section{Introduction}
\label{sec:introduction}
\IEEEPARstart{B}{lockchain} has emerged as a critical technology to support distributed applications~\cite{bitcoin, eth}. 
However, traditional blockchain systems~\cite{bitcoin, eth} fall short regarding scalability.
Sharding is a promising approach to improving blockchain scalability \cite{omniledger,rapidchain, li2025sp}.
It partitions nodes in the blockchain network into multiple disjoint committees (shards).
Each shard maintains a subset of the whole blockchain ledger state and conducts intra-shard consensus to generate blocks and process transactions in parallel.
In general, under the same network scale, configuring \emph{a large amount of small shards helps to achieve higher transaction concurrency} and throughput (i.e., transactions per second, TPS) \cite{divideAndScale, cochain, gearbox}.

However, naively configuring numerous small shards can easily lead to certain shards becoming corrupted, compromising security. To ensure security, existing permissionless blockchain sharding systems typically adopt relatively large shard sizes \cite{chainspace, pyramid, omniledger, elastico, rapidchain, cycledger, li2025sp}. Under the same network scale, such \emph{larger shard sizes significantly reduce the total number of shards in the network and diminish the efficiency of intra-shard consensus, consequently severely restricting the transaction concurrency} (TPS) of sharding-based blockchain systems.
Specifically, nodes within blockchain sharding systems (especially permissionless ones) are typically allocated to shards randomly \cite{elastico, rapidchain, monoxide, li2022jenga, LBCHAIN, li2025sp}. \emph{Due to this randomness}, when the network size and overall malicious node ratio remain unchanged, naively adopting small shard sizes can easily lead to excessively high malicious node ratios in some shards, causing shard corruption (e.g., $\geq1/3$ malicious nodes for BFT-type (Byzantine Fault Tolerance) consensus under partial-synchronous network \cite{castro1999practical}) and compromising system security \cite{bitcoin}.
Therefore, most sharding systems have to \emph{set large shard size to ensure a negligible failure probability for each shard}, guaranteeing security.
For example, in OmniLedger~\cite{omniledger}, a shard could include $600$ nodes when 1/4 of the network nodes are malicious. 

In this paper, we aim to boost transaction concurrency for blockchain sharding securely and efficiently.
This is achieved by proposing a blockchain sharding system named SpiralShard.
Unlike most previous works, which require each shard to be uncorrupted, our intuition is to allow the existence of \emph{some} corrupted shards with a larger fraction of malicious nodes. 
Therefore, the system has more and smaller shards for higher concurrency.
To prevent corrupted shards from compromising system security, we design an \ac{LCE} protocol.
Specifically, shards are divided into multiple \emph{endorsement groups}, each belonging to one group. 
Each shard must include block headers from other shards within its blocks and \emph{reach intra-shard consensus to express its endorsement decision}.
Blocks within a shard require endorsement in sequence from all other shards within the same endorsement group to be finalized securely.
With the help of endorsement groups, each shard can securely finalize its blocks, even if there are forks.
As a result, the shard size is reduced without violating security, and the concurrency is boosted.

It is not straightforward to design the \ac{LCE} protocol. We summarize the main challenges as follows.

\vspace{3pt}
\noindent
1). \textbf{How to maintain low overheads for each shard during cross-shard verification?}
The LCE protocol leverages cross-shard verification and endorsement to secure corrupted shards, introducing additional overhead.
To validate blocks from other shards, a straightforward idea is to store the state of other shards and verify raw transactions, leading to higher storage, computation, and communication overhead for each node.
Hence, we need to reduce the overhead of cross-shard verification and endorsement.

To reduce overhead, our core idea is that \emph{each shard internally ensures the validity of individual transactions through consensus} (preventing transaction manipulation), and \emph{subsequently employs the LCE protocol to address potential fork problems. }
Consequently, cross-shard verification of individual transactions is unnecessary; instead, shards \emph{only need to exchange lightweight block headers to detect forks.} 
To realize this goal, we rigorously conduct probabilistic modeling and derive explicit mathematical formulations (in Section \ref{epoch_security}). 
Based on this modeling, we precisely fine-tune shard sizes to ensure—except with negligible failure probability (formally defined in Section \ref{subsec:obj})—that the \emph{proportion of malicious nodes within each shard is always below 2/3}. 
This threshold is selected because our analysis shows that when the fraction of malicious nodes is below the quorum size (i.e., 2/3 for BFT consensus protocols in partially synchronous networks) required for consensus, invalid transactions cannot accumulate enough malicious signatures to pass the consensus procedure. 
Hence, under such conditions, each shard inherently guarantees transaction validity via internal consensus. 
However, corrupted shards may still produce forks that cannot be resolved independently. 
In this scenario, the LCE protocol efficiently resolves forks solely by monitoring block headers, thereby maintaining low overhead.

\begin{figure}[t]
\centerline{\includegraphics[width=0.33\textwidth]{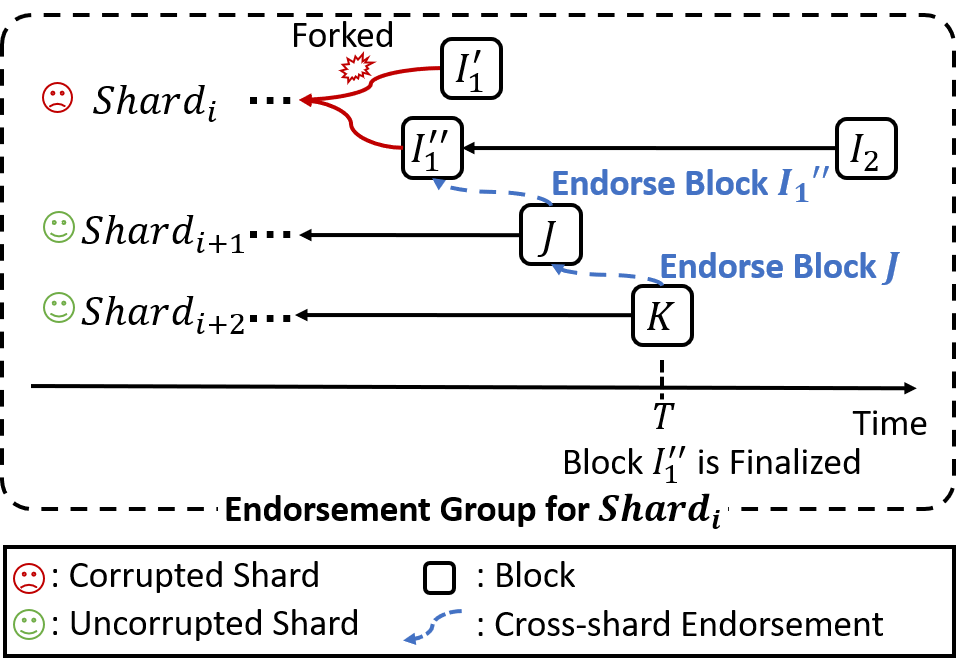}}
\vspace{-9pt}
\caption{Linked Cross-shard Endorsement for $Shard_i$.}
\label{fig:OCE}
\end{figure}

\vspace{3pt}
\noindent
2). \textbf{How to preserve safety and liveness in LCE protocol while some shards are corrupted?}
As discussed above, some randomly corrupted shards may threaten system \textbf{\emph{safety}} (i.e., honest nodes agree on the same and valid shard chain for each shard) by creating forks. 
To securely address the fork issue, we first require \emph{multiple shards to form an endorsement group}, within which shards perform mutual verification and endorsement. 
More importantly, through rigorous probabilistic analysis, we carefully adjust the size of each endorsement group, enabling the system to include as many endorsement groups as possible while simultaneously ensuring (except with negligible failure probability) that \emph{each endorsement group contains at least one honest shard}. 
Consequently, honest shards within an endorsement group guarantee the honest selection of only one single fork for endorsement. This mechanism ensures that no shard can have multiple conflicting forks being finalized, as no shard can have multiple forks endorsed by all members of its endorsement group.

However, the aforementioned design could result in different honest shards within an endorsement group endorsing distinct forks. Consequently, no single fork would obtain endorsements from all group members, preventing any fork from being finalized and thereby impairing system \textbf{\emph{liveness}} (the system does not stall indefinitely). 
To address this issue, we enforce a predetermined endorsement order (e.g., ascending order of the shard IDs) among the shards within each endorsement group, whereby \emph{each shard endorses only the decision previously made by its predecessor shard}. 
Consequently, the endorsement from each shard implicitly validates the endorsements provided by all preceding shards. 
Under this mechanism, each shard's consensus outcome—reflected in the block header—accumulates a \emph{sequential chain of linked endorsements from the endorsement group members}. 
This sequentially linked endorsement strategy ensures that honest shards invariably endorse the same fork consecutively, enabling a single fork to receive unanimous endorsements from all endorsement group members and thereby achieve finality.

For example, in Figure~\ref{fig:OCE}, corrupted $Shard_{i}$ is forked and is required to eliminate the fork with the help of shards in the endorsement group. 
Without loss of generality, the endorsement group consists of $Shard_{i}$, $Shard_{i+1}$, and $Shard_{i+2}$.
$Shard_{i}$ is forked and produces blocks ${I}_{1}^{'}$ and ${I}_{1}^{''}$ via intra-shard consensus.
These blocks can be inherently considered as receiving $Shard_{i}$'s endorsement since passing the intra-shard consensus.
However, only block ${I}_{1}^{''}$ is sequentially endorsed by $Shard_{i+1}$'s block $J$.
Consequently, block $J$ is endorsed by $Shard_{i+2}$'s block $K$, thus original block ${I}_{1}^{''}$ is endorsed by all shards in the endorsement group.
Hence, only block ${I}_{1}^{''}$ is finalized.
In this case, the fork in $Shard_{i}$ is eliminated, and $Shard_{i}$ has to extend the chain after block ${I}_{1}^{''}$.


To further enhance system liveness against possible liveness attacks by corrupted shards (i.e., hindering the advancement of the intra-shard consensus or LCE protocol),
we adopt a \emph{mixed threat model} comprising both \emph{Byzantine nodes}, which simultaneously compromise safety and liveness, and \emph{alive-but-corrupt (a-b-c) nodes} \cite{flexible} (CCS '19), which maintain liveness but exclusively attack safety.
Under this mixed threat model, we ensure that the fraction of Byzantine nodes in each shard remains below $1/3$, and we allow corrupted shards to contain \emph{additional} no more than $1/3$ of a-b-c nodes.
These measures are also guided by our rigorous theoretical analysis and shard size fine-tuning.
In this case, the continued advancement of the intra-shard consensus and LCE protocol is guaranteed, refining liveness.

\vspace{3pt}
\noindent
3). \textbf{How to maintain high efficiency when there are some slow shards in the system?}
In SpiralShard, we \emph{pipeline} the block generation to enhance overall throughput.
Generally, shards may have different block generation speeds.
Meanwhile, the \emph{sequentially processed cross-shard endorsement may be halted} due to slow shards.
To solve this problem, we design a pipelining mechanism to allow each shard to generate new blocks \emph{optimistically}.
Therefore, each shard can generate new blocks before its blocks are finalized.
To ensure the \emph{security} of the proposed pipelining mechanism, we require that the generated blocks be finalized only after being endorsed by the whole endorsement group.
To preserve efficiency, our pipelining mechanism requires a shard to endorse the \emph{latest block} instead of every block of the preceding shard. 
With this design, when a block is endorsed by all shards from its group, \emph{this block and its parent blocks are all finalized}.

\vspace{3pt}
This paper mainly has the following contributions:
\begin{itemize}[left=0pt]
  \item We propose SpiralShard, a blockchain sharding system that improves transaction concurrency by allowing some shards to be corrupted by larger fractions of malicious nodes.
  \item We design the LCE protocol that provides higher resiliency within shards. It offers an additional tolerance for no larger than $1/3$ a-b-c nodes based on less than $1/3$ fraction of Byzantine nodes.
  \item We implement SpiralShard based on Harmony~\cite{Harmony} and conduct evaluations based on large-scale Amazon EC2 deployment. 
  Compared with Harmony, SpiralShard achieves around 19$\times$ throughput under a large network size with 4,000+ nodes.
\end{itemize}

%% file: Files/relatedWork_motivation.tex
\section{Related Work and Motivation}

\subsection{Blockchain and Blockchain Sharding}
Traditional blockchain systems require nodes to store and verify every transaction on a single chain, resulting in poor performance and scalability.
For example, Bitcoin can only process 7 tx/s (transaction per second), while Ethereum can only process 15 tx/s~\cite{huang2021survey}.
Recent works adopt sharding \cite{elastico} to improve the scalability of blockchain~\cite{elastico,omniledger,rapidchain,chainspace,sgxSharding, monoxide, li2022jenga, LightCross, accountMigra, li2025sp}.
The main idea of sharding is to partition the nodes into multiple small committees (shards).
Each shard maintains a disjoint subset of the whole blockchain state, conducts intra-shard consensus, and processes distinct transactions in parallel.
However, existing blockchain sharding systems still suffer from \emph{limited concurrency.}

\subsection{Shard Size and Concurrency}
\label{subsec:Related_size}

We find that the main reason for the poor concurrency of the existing blockchain sharding systems is the \emph{large shard size}, resulting in slower intra-shard consensus and fewer shards.
However, large shard sizes are set to limit the fraction of malicious nodes in \emph{each shard} under its fault tolerance threshold when the nodes are randomly assigned.

Some works \cite{elastico, omniledger, rapidchain, sgxSharding, pyramid, cycledger} try to improve system concurrency by reducing the shard size. 
However, they have various limitations.
For instance, some systems \cite{rapidchain,cycledger, li2025sp} tolerate a larger fraction (i.e., $<1/2$) of malicious nodes within each shard based on a less practical assumption that the network within each shard is synchronous. 
However, real-world network environments are often not synchronous (e.g., partial-synchronous), especially for large-scale blockchain systems.
SGX-Sharding \cite{sgxSharding} reduces the shard size by applying trusted hardware. 
However, it brings additional deployment overhead to each node, limiting its generality.
Some other solutions reduce the shard size at the cost of reducing the systems' overall resiliency, compromising the system's security.
For example, in Pyramid~\cite{pyramid}, the assumption entails a smaller fraction of malicious nodes in the network, resulting in a system capable of withstanding less than $1/8$ of malicious nodes.

\vspace{3pt}
\noindent
\textbf{Motivation and Intuition.}
Can we safely configure small shard sizes, enhancing the transaction concurrency?
Our intuition is to allow some shards to be corrupted and leverage honest shards to recover corrupted shards from forking.

\subsection{Blockchain Sharding with Corrupted Shards}

Only a handful of blockchain sharding studies allow corrupted shards, with various limitations.
Free2Shard~\cite{rana2022free2shard} allows corrupted shards and preserves the system's security via a network-wide consensus.
However, it is based on the assumption of a global synchronous network environment. 
Therefore, it does not work correctly under a practical partial-synchronous network. 
In addition, Free2Shard adopts the \ac{PoW} consensus protocol, which does not provide deterministic finality and is prone to forking problems.
PolyShard \cite{li2020polyshard} leverages coded sharding to enable nodes to store and verify coded shard data, thereby maintaining system security even in the presence of corrupted shards. 
However, PolyShard also requires an ideal synchronous network environment, which may limit its practicality. 
Additionally, the computational overhead of polynomial encoding/decoding and the communication costs of network-wide data exchange could impact both its performance and generality.
In GearBox~\cite{gearbox}, the system relies on a network-wide consensus to ensure the system's security, which can easily lead to a performance bottleneck in large-scale networks.
The latest work, known as CoChain~\cite{cochain}, similarly stems from an intuition that allows shard corruption to enhance concurrency.
In CoChain, shards are randomly assigned to CoC (Consensus on Consensus) groups.
However, CoChain only permits the failure of less than one-third of the shards within a group. 
When the group size is no larger than 3, CoChain cannot allow any corrupted shard, losing feasibility.
More importantly, CoChain needs to rely on very time-consuming state migration to replace the corrupted shards when they fork. 
In contrast, our SpiralShard can utilize the LCE protocol to easily help corrupted shards remove excess forks without the need for shard replacement or recovery.


\vspace{3pt}
SpiralShard is a novel design that is different from existing blockchain sharding systems. 
Built upon the assumption of partial-synchronous networks, it is more practical. 
Leveraging BFT-typed intra-shard consensus ensures finality and energy efficiency. 
Unlike systems reliant on global consensus to resolve forks, it leverages multiple endorsement groups, preserving scalability and low overhead.
Through a series of deliberate design choices--including the LCE protocol and the controlled tolerance of statistically bounded corrupted shards--SpiralShard raises transaction-level concurrency while rigorously preserving system security. Table \ref{table:qualitative-comparison} provides a qualitative comparison between SpiralShard and several representative blockchain sharding schemes.

\begin{table*}[ht] 
\caption{Qualitative comparison of representative blockchain sharding protocols.}
\centering{}%
\begin{tabular}{c|c|c|c|c|c|c}
\hline
\hline 
\multirow{2}{*}{Protocol} & \multirow{2}{*}{Corrupted Shards} & \multirow{2}{*}{Network Model} & \multirow{2}{*}{Consensus Type} & Network-wide & \multirow{2}{*}{Protocol Overhead} & \multirow{2}{*}{Concurrency} \tabularnewline
 & & & & Communication & & \tabularnewline
\hline 
\hline
\cite{elastico, omniledger, pyramid, li2022jenga}, etc. & Not Allowed & Partial-synchronous & BFT-type & Needless & Low & Low \tabularnewline
\hline 
\cite{rapidchain, cycledger, repchain, li2025sp}, etc. & Not Allowed & Synchronous & BFT-type & Needless & Low & Medium \tabularnewline
\hline 
Free2Shard \cite{rana2022free2shard}, PolyShard \cite{li2020polyshard} & Allowed & Synchronous & PoW & Need & High & Medium \tabularnewline
\hline 
GearBox \cite{gearbox} & Allowed & Partial-synchronous & BFT-type & Need & High & Medium \tabularnewline
\hline 
\textbf{\emph{SpiralShard}} & \textbf{\emph{Allowed}} & \textbf{\emph{Partial-synchronous}} & \textbf{\emph{BFT-type}} & \textbf{\emph{Needless}} & \textbf{\emph{Low}} & \textbf{\emph{High}} \tabularnewline
\hline 
\hline
\end{tabular}
\label{table:qualitative-comparison}
\end{table*}

%% file: Files/model.tex
\section{Network and Threat Model}

 \subsection{Network Model}
 \label{subsec_network_model}
 
SpiralShard works on a \emph{partial-synchronous} \ac{P2P} network, in which there is a known bound $\Delta$ and an unknown Global Stabilization Time (GST).
After GST, the network becomes synchronous, and all transmissions between two honest nodes arrive within time $\Delta$.
Like most previous blockchain sharding systems, the messages are propagated through gossip protocol.
Each node in the system has its public/secret key pair, which can be authenticated through a trusted Public-Key Infrastructure (PKI) to represent its identity while sending messages.
Like many prior works~\cite{pyramid, omniledger, Harmony}, SpiralShard utilize a beacon chain
to publicly store the identities of participants.

In SpiralShard, the network consists of $N$ nodes. 
The shard size (i.e., the number of nodes per shard) and the endorsement group size (i.e., the number of shards per endorsement group) are denoted as $S$ and $G$, respectively. 
This implies that each endorsement group contains $S \cdot G$ nodes, with $\lfloor N/S \rfloor$ shards within the network. 
A node's endorsement group identity is determined by its shard identity (i.e., $GroupID = \lfloor ShardID / G \rfloor$).

SpiralShard employs the account/balance model, widely adopted by existing works to represent ledger states~\cite{pyramid, Harmony, monoxide, LBCHAIN, sgxSharding, sharper, brokerchain}. 
In SpiralShard, a specific account's state (e.g., balance) is maintained by one shard, determined by the hash of its account address. 
Accordingly, a transaction in the network is routed to the corresponding shard based on its associated account address.
It is important to emphasize that this includes both intra-shard and cross-shard transactions.
Besides, the accounts involved in cross-shard transactions may be in different endorsement groups.

\subsection{Threat Model}
\label{subsec:threat_model}
In SpiralShard, there are honest and malicious nodes.
The honest nodes obey all the protocols.
There are two kinds of malicious nodes: \textbf{\emph{Byzantine}} and \textbf{\emph{alive-but-corrupt (a-b-c)}} nodes \cite{flexible}.
The Byzantine nodes may collude together (both intra-/cross-shard) and corrupt the system arbitrarily, such as remaining silent, sending conflicting messages to different nodes, and voting for blocks containing invalid transactions.
Besides, bounded by the eventual synchrony property of the partial-synchronous network, Byzantine nodes can only exploit short-term message delays or reorderings. 
A-b-c nodes, proposed by Flexible BFT\cite{flexible} (CCS '19), are widely recognized as a typical and rational type of attacker, especially in real-world permissionless blockchains~\cite{jo2020toward, kane2021highway, Trebiz, flexible}.
A-b-c nodes always properly attempt to sign and send inconsistent messages to different nodes to break safety (e.g., generate fork chains).
However, they \emph{positively vote for proposals to push consensus forward (i.e., maintain liveness)}.
Align with the previous work~\cite{flexible}, \emph{a-b-c nodes should be treated as Byzantine during the safety proof.
And once we have proved safety, we can treat a-b-c nodes as honest while proving liveness. }

\vspace{3pt}
\noindent
\textbf{Justification of A-b-c Nodes.}
A-b-c nodes are considered a typical and rational type of attacker~\cite{jo2020toward, kane2021highway, Trebiz, flexible}, especially in permissionless blockchain systems.
The rationale behind this stems from the nature of blockchain payment systems, wherein some attackers often benefit by compromising system safety, evidenced by the substantial losses incurred due to double-spending attacks in Bitcoin and Ethereum Classic~\cite{etcdouble}. 
On the other hand, nodes are motivated to preserve liveness to earn service fees~\cite{jo2020toward, kane2021highway,ranchal2020zlb, Trebiz, blockchainIsDead, basilic, flexible}.

\vspace{3pt}
SpiralShard is a blockchain sharding system that employs a BFT-typed consensus mechanism within each shard. Operating under a partially synchronous network model, SpiralShard has an overall fault tolerance threshold of 1/3. 
Specifically, to maintain system security, the total fraction of malicious nodes $F$ (including Byzantine and a-b-c nodes) across the entire network must remain below 1/3. In other words, the total number of malicious nodes are $N \cdot F$.
The fraction of Byzantine and a-b-c nodes in the system are denoted as $F_B$ and $F_A$, and $F_B+F_A=F$.
Correspondingly, there are $N \cdot F_B$ Byzantine and $N \cdot F_A$ a-b-c nodes in the system, respectively.

\vspace{3pt}
\noindent
\textbf{Slowly-Adaptive Adversaries.}
Like most blockchain systems \cite{rapidchain, omniledger, li2022jenga, repchain, sgxSharding, pyramid, brokerchain}, the adversaries in the system are slowly-adaptive, i.e., the set of malicious nodes and honest nodes in each shard are fixed during each epoch and can only be changed after shard reconfiguration~\cite{buildingBlocks_survey} (see Section \ref{subsec:other}).
In other words, malicious nodes may attempt to corrupt honest nodes (e.g., through bribery or other means). However, such adaptive attacks require at least one epoch to take effect. Due to the randomized shard reconfiguration process conducted at each epoch transition, these slowly adaptive malicious nodes cannot compromise this randomness, thereby preventing the aggregation of an excessive number of malicious nodes within any particular shard.

%% file: Files/system_overview.tex
\section{System Overview and Objectives}

\subsection{Overview of Main Designs}

Existing large-scale permissionless blockchain sharding systems typically require each shard to be honest (e.g., maintaining $<1/3$ fraction of malicious nodes within each shard for BFT-type consensus under partially synchronous networks) to guarantee the security of the entire system. 
However, from a probabilistic perspective, due to the random distribution of nodes into shards, ensuring that every shard remains honest except with negligible failure probability necessitates configuring excessively large shard sizes. 
Such oversized shard configurations inevitably reduce the transaction concurrency of the system.

To enhance system performance, SpiralShard first allows the existence of some corrupted shards (where $\geq 1/3$ fraction of malicious nodes within a shard). Consequently, under the same network size and overall proportion of malicious nodes, it becomes feasible to configure more and smaller shards, thus significantly improving transaction concurrency.
Second and more importantly, to simultaneously boost transaction concurrency and maintain security, we innovatively propose the Linked Cross-shard Endorsement (LCE) protocol. LCE acts as a secondary security protocol operating across shards on top of intra-shard consensus. Its core principle is leveraging the collaborative effort of multiple shards to recover corrupted shards and safeguard overall system security. 

\begin{figure}[t]
\centerline{\includegraphics[width=0.42\textwidth]{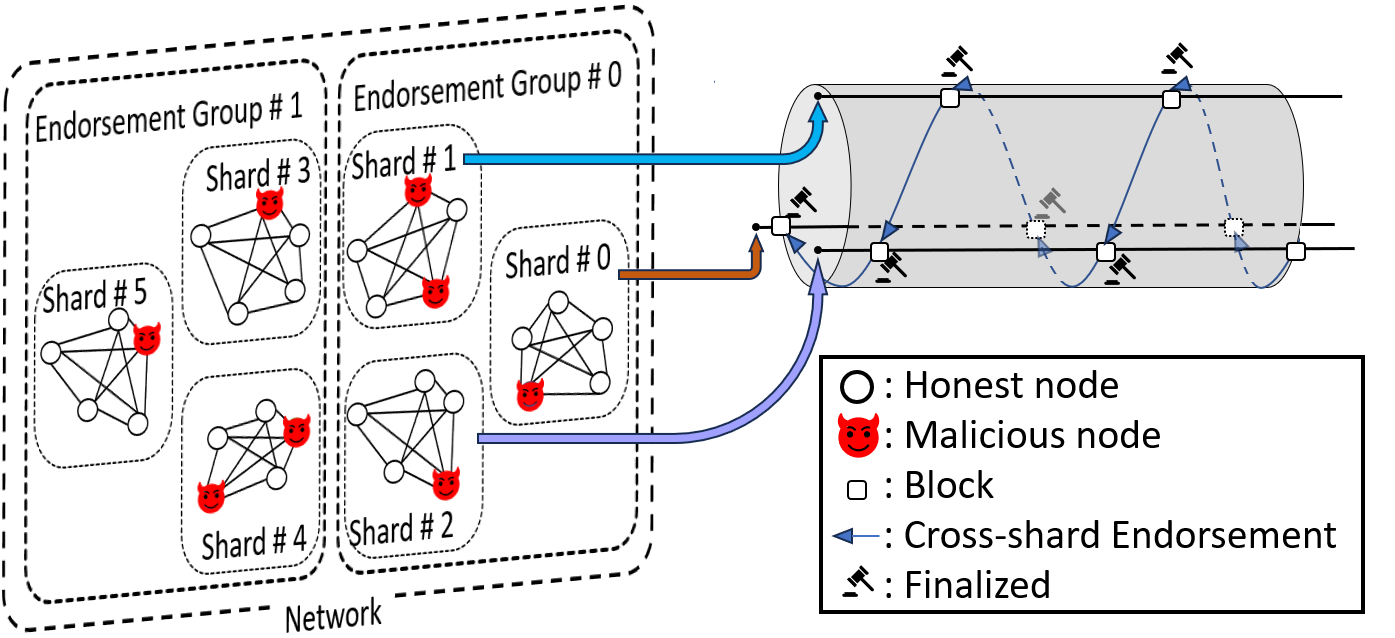}}
%
\vspace{-9pt}
\caption{System architecture.} 
\label{fig:system_architecture}
\end{figure}

Specifically, as illustrated in Figure \ref{fig:system_architecture}, multiple shards, including at least one honest shard, constitute an endorsement group, collectively securing each shard within the group. 
To ensure (except with negligible failure probability) that each endorsement group contains at least one honest shard, we calculate appropriate shard sizes and endorsement group sizes via rigorous probabilistic modeling and mathematical derivation (Section \ref{epoch_security}).
Multiple such endorsement groups coexist within the system simultaneously.
Shards within an endorsement group individually run their intra-shard consensus and sequentially performs cross-shard endorsements on the endorsement results from its preceding shard, strictly adhering to a predetermined order. 
Consequently, any intra-shard consensus outcome obtains sequentially linked endorsements. 
Once a shard’s consensus outcome has been sequentially and fully endorsed by all shards in its endorsement group, it can be considered finalized and secure (Section \ref{subsec:basicOCE}). 
This is primarily because the presence of the honest shard within the endorsement group ensures it will neither engage in malicious behavior nor endorse fraudulent endorsements.

Furthermore, SpiralShard employs a mixed threat model and precisely adjusts the shard sizes (again, through rigorous probabilistic derivations (Section \ref{epoch_security})) to ensure, except with negligible failure probability, that each shard's fraction of malicious nodes remains below 2/3 (as opposed to the conventional threshold of 1/3). This includes less than 1/3 Byzantine nodes and up to 1/3 alive-but-corrupt (a-b-c) nodes. This approach provides two primary benefits.
First, the mixed threat model enhances system liveness because a-b-c nodes solely target safety, not liveness. Second and more importantly, limiting the proportion of malicious nodes within each shard to below 2/3 significantly reduces the overhead associated with cross-shard endorsement. At this threshold, adversaries cannot control enough malicious nodes within a shard to let invalid transactions pass intra-shard consensus, thus ensuring transaction validity internally within each shard. Nonetheless, adversaries may still execute equivocation attacks to cause forks. Fortunately, such forks can be efficiently detected using lightweight block headers. Consequently, the LCE protocol enables the transmission of only block headers (rather than entire blocks) as intra-shard consensus results for cross-shard communication. As a result, the LCE protocol can effectively resolve intra-shard forks through connected cross-shard endorsements at minimal overhead, thereby securing the overall system (Section \ref{subsub:efficient_communication}).

\subsection{Objectives}
\label{subsec:obj}

SpiralShard aims to achieve the following objectives:

\begin{itemize}[left=0pt]
    \item \textbf{Security}: Given that the total fraction of malicious nodes $F$ within the entire system is below its fault tolerance threshold (see Section \ref{subsec:threat_model}), SpiralShard aims to guarantee system security (i.e., safety and liveness) \emph{except with negligible failure probability}. 
    {\color{red}
    In this paper, we define the negligible failure probability as $<2^{-17}$, indicating the system is expected to fail only once every 359 years for one-day epoch.
    This definition is consistent with many existing blockchain sharding systems \cite{pyramid, li2022jenga, rapidchain, omniledger, cochain}.
    }
    Specifically, for each shard, the system ensures—except with negligible failure probability—that it contains fewer than 2/3 malicious nodes, including fewer than 1/3 Byzantine nodes and at most 1/3 a-b-c nodes. 
    For each endorsement group, the design ensures—except with negligible failure probability—the inclusion of at least one honest shard (i.e., a shard with a malicious node proportion below 1/3). 
    After achieving these two objectives, the Linked Cross-shard Endorsement (LCE) protocol further aims to guarantee both safety (i.e., honest nodes agree on the sequence of the same and valid finalized blocks in each shard chain \cite{rapidchain, omniledger}) and liveness (i.e., transactions in each shard are finalized within finite time, the system does not stall indefinitely \cite{rapidchain, omniledger}).

    \item \textbf{High Concurrency}: By securely allowing shards to contain a higher fraction of malicious nodes ($<2/3$ rather than the conventional $<1/3$), SpiralShard aims to configure a greater number of smaller-sized shards within the same network scale and overall malicious node ratio. On one hand, this increases the total number of shards in the system; on the other, it enhances the efficiency of intra-shard consensus. Through these combined effects, SpiralShard aims to significantly boost overall system throughput (transactions per second, TPS), achieving higher transaction concurrency.
\end{itemize}

The proofs, analysis, and experiments related to the aforementioned objectives are detailed in Section \ref{sec:analysis} and Section \ref{sec:evaluation}.

\subsection{Overview of Other Components}
\label{subsec:other}
Beyond the overview of our main designs, we now provide brief remarks about other notable components in the system.



\vspace{3pt}
\noindent
\textbf{Epochs.}
SpiralShard operates based on epochs of fixed duration. The exact length of an epoch may vary according to specific system implementations, but without loss of generality, we set the epoch length to one day, consistent with many existing systems \cite {sgxSharding, pyramid, rapidchain}. Within each epoch, every shard continuously runs its intra-shard consensus to advance the system's state. More importantly, each shard in SpiralShard simultaneously executes the LCE protocol to ensure system security. 

\vspace{3pt}
\noindent
\textbf{Shard Reconfiguration.}
Between epochs, SpiralShard performs standard shard reconfiguration (consistent with most existing blockchain sharding systems \cite{rapidchain, pyramid, cycledger, Harmony}) to reassign nodes among shards. 
This procedure primarily aims to defend against slowly-adaptive adversaries, preventing them from gradually corrupting individual shards.
For security, the nodes are randomly assigned to different shards.
Specifically, a random seed is generated at the end of each epoch (e.g., by a decentralized randomness protocol or via the beacon chain, like RANDAO in Ethereum 2.0 or other ways \cite{eth, Harmony}). 
Similar to many existing sharding systems \cite{omniledger, rapidchain, pyramid, li2022jenga, Harmony}, using this seed, a VRF (Verifiable Random Function) \cite{micali1999verifiable} is run by each node to determine if it is selected to randomly move to a different shard (or to remain). Additionally, a VDF (Verifiable Delay Function) \cite{boneh2018verifiable} can be used to make the randomness unpredictable until it is time to apply it, to prevent adversaries from precomputing favorable placements. In this way, we can guarantee that the node reassignment of each shard during the shard reconfiguration is random, verifiable, unbiased, and unpredictable. Due to this randomness, the adversary cannot ensure their nodes stay together in one shard after reconfiguration. 


\vspace{3pt}
\noindent
\textbf{Cross-shard Transactions.}
SpiralShard can leverage existing mechanisms to handle \emph{cross-shard transactions}, even when these transactions span across different endorsement groups.
However, in SpiralShard, to ensure the security of cross-shard transactions, we require that cross-shard transactions be \emph{processed and finalized in other shards only after they are finalized by the LCE protocol.}
Specifically, the cross-shard \emph{transfer} transactions are processed via the existing relay-based mechanism~\cite{Harmony, monoxide}.
Moreover, our system is inherently capable of supporting \emph{smart contracts}, primarily because we ensure the safe finalization of blocks. 
To facilitate cross-shard contract transactions, we can leverage the traditional Two-Phase Commit (2PC) protocol \cite{sgxSharding, li2022jenga}. 
The detailed discussion about the cross-shard smart contract processing and its limitation are given in Section \ref{subsec:cx_contract}.

%% file: Files/protocol_design.tex
\section{System design}

We first introduce the basic LCE protocol in Section \ref{subsec:basicOCE}. Building upon this, Section \ref{subsub:efficient_communication} describes how we leverage block headers to reduce the overhead associated with cross-shard endorsements. Finally, Section \ref{subsec:pipe} presents a pipelining mechanism designed to safely and further enhance the protocol's performance.

\begin{figure*}[t]
\centerline{\includegraphics[width=0.81\textwidth]{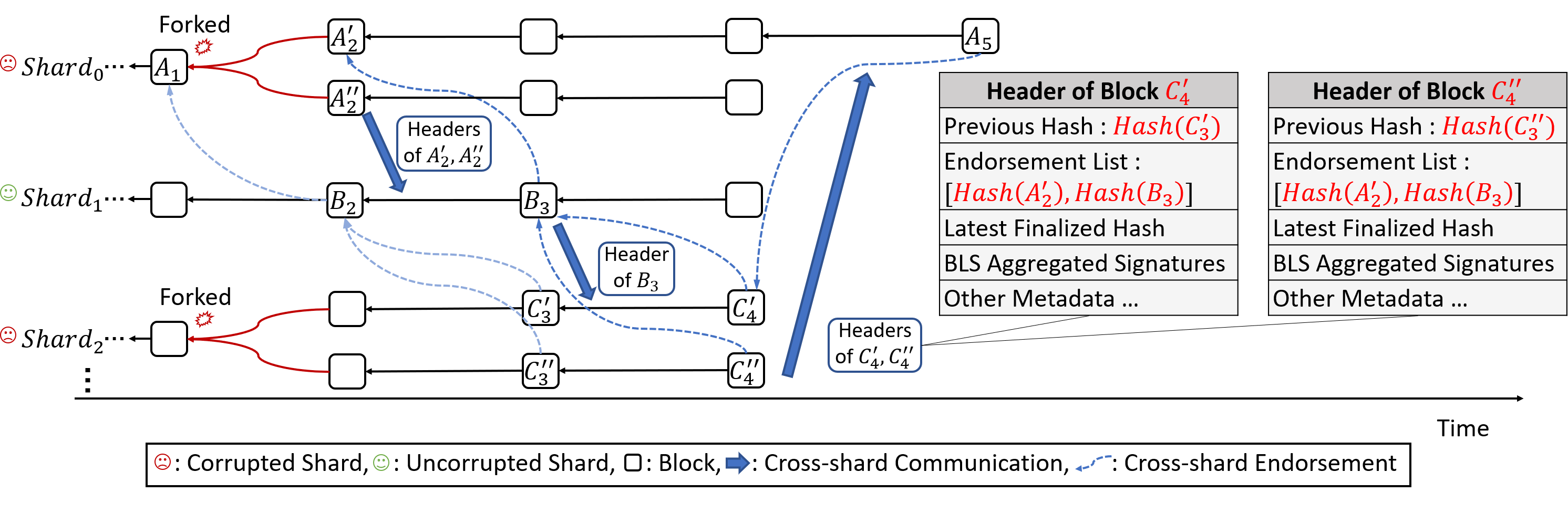}}
\vspace{-18pt}
\caption{Linked Cross-shard Endorsement for blocks $A_2'$.}
\label{fig1}
\end{figure*}
 
\subsection{Basic Design of LCE}
\label{subsec:basicOCE}

The LCE protocol requires that after a block is appended to its shard chain via intra-shard consensus, it must be sequentially endorsed by all shards (via each shard's intra-shard consensus) in its endorsement group before being finalized.
To illustrate the process of a block, from being appended to its shard chain to being finalized, the \ac{LCE} protocol declares three states for each block: 1) prepared, 2) finalized, and 3) discarded. 
A block that passes the intra-shard consensus is considered prepared and would be finalized if it is endorsed by all shards from its endorsement group later.
On the other hand, a prepared block that conflicts with finalized ones is considered discarded.

The basic design requires transmitting blocks across shards since they contain the history of each step of linked cross-shard endorsement.
However, in our final design, only transmitting block headers is sufficient, reducing the cross-shard communication overhead.
We will illustrate this later in Section~\ref{subsub:efficient_communication}.
The procedure of \ac{LCE} includes three phases:

\vspace{3pt}
\noindent
\textbf{Block Preparation}.
In SpiralShard, each shard runs \textbf{\emph{BFT-typed intra-shard consensus}} to append blocks to its shard chain.
During the consensus, nodes can verify whether a block has passed the intra-shard consensus through the \emph{signatures in the block header}.
Such a block that passes the intra-shard consensus is valid (does not conflict with the parent block) and considered \emph{prepared} and \emph{receives the endorsement from its shard}.
Honest nodes must broadcast the prepared block they vote for to the next shard for cross-shard endorsement.
An honest shard always appends only one prepared block to the tail of its shard chain each time.
However, when a shard is corrupted, it can append multiple prepared blocks after the same block, resulting in a fork.

As shown in Figure~\ref{fig1}, blocks $A_2'$ and $A_2''$ are prepared blocks, forming a fork in $shard_0$.
Once the blocks are prepared, nodes send the block they vote for to $Shard_1$ for being endorsed.
Then, we leverage \emph{Cross-shard Endorsement and Block Finalization} to eliminate the fork in $Shard_0$.

\vspace{3pt}
\noindent
\emph{Choice of Intra-shard Consensus.} 
The intra-shard consensus protocol is a pluggable component in SpiralShard, as long as it is a standard BFT-typed consensus and can operate in a partial-synchronous network (it uses a $2/3$ quorum size and tolerates less than $1/3$ Byzantine nodes, like \ac{PBFT}~\cite{castro1999practical}).
In this work, we choose the \ac{FBFT}~\cite{Harmony} consensus protocol proposed by Harmony \cite{Harmony} as our intra-shard consensus protocol. 
\ac{FBFT} is a variant of \ac{PBFT}~\cite{castro1999practical}, adopting a short signature scheme, \ac{BLS} multi-signature~\cite{bls} to collect votes from participants for higher scalability.

\ac{FBFT} has a view change mechanism
to replace the Byzantine leader of consensus, similar to \ac{PBFT}~\cite{castro1999practical}.
Moreover, like existing BFT-type consensus protocols in partially synchronous networks, FBFT also incorporates a timeout mechanism to trigger the view change process for liveness. 
Similarly, during the asynchronous phase before GST, the timeout duration gradually increases with each view change and eventually converges to a stable value (once the network becomes synchronous), to accommodate the asynchronous nature of the network and prevent overly frequent view changes.

{\color{red}
After GST, the timeout in SpiralShard converges to a small constant value (on the order of only a few seconds). 
This behavior is consistent with standard partially synchronous consensus protocols like PBFT and Tendermint \cite{castro1999practical, buchman2016tendermint}, and we inherit it in SpiralShard’s implementation (using FBFT \cite{Harmony}) to ensure timely view changes with minimal cost.
As a result, any leader replacement (view change) after GST completes rapidly, incurring minimal overhead. 
}
Additionally, under normal conditions, FBFT regularly rotates leaders (e.g., once per epoch) based on the identity of each consensus node, aligning with many existing BFT-type consensus protocols \cite{buchman2016tendermint, Harmony}. Since we adopt a leader rotation approach similar to existing BFT-type consensus mechanisms and further ensure the overall system's security through the LCE protocol, the leader replacement process in SpiralShard is also secure. A detailed proof is presented in Section \ref{protocol_security}.


\vspace{3pt}
\noindent
\textbf{Cross-shard Endorsement}.
The LCE protocol utilizes the endorsement group of multiple shards to decide on one of the fork branches on a corrupted shard.
Since an honest shard will only endorse blocks on one fork branch on a corrupted shard to eliminate forks, a group must contain at least one honest shard except with negligible failure probability.
To ensure that multiple honest shards within the same endorsement group make consistent endorsement decisions, the LCE protocol requires that shards conduct endorsements based on the ascending order of Shard IDs.

In this phase, each shard is responsible for monitoring the previous shard and making endorsement decisions via intra-shard consensus.
Specifically, it verifies the validity of the prepared block received from the preceding shard and selects a prepared one.
Then, it conducts intra-shard consensus to record this \emph{prepared} block's hash into its block header. 
Each honest node will ensure that the endorsement decision in this block does not conflict with that in its parent blocks (i.e., this block and its parent blocks endorse on the same fork branch from the preceding shard). 
Once such a block passes intra-shard consensus, it records the endorsement decision for its shard.

\vspace{3pt}
\noindent
\emph{Fork Selection.} 
In the LCE protocol, the choice of which fork to endorse is crucial, as it determines which branch within a corrupted shard will ultimately be finalized by the LCE protocol. In a partially synchronous network, deterministically identifying the correct fork is impossible. Fortunately, SpiralShard does not require deterministic fork detection. In the LCE protocol, the leader follows two main rules when selecting a fork from the previous shard.
First, the selected fork must be an extension of the previously finalized fork as determined by the LCE protocol, to prevent conflicts. Second, there may be multiple forks that extend from the same finalized fork. Among these forks, the leader will select the longest one from its own perspective, and then propose the latest block on that fork for endorsement along with other transactions through intra-shard consensus.
For the honest nodes in the shard, they verify during consensus whether the fork chosen by the leader is indeed an extension of the previously finalized fork and whether the leader selected only one fork for endorsement (to prevent equivocation attacks). However, the honest nodes do not verify if the leader has chosen the longest fork, since the information observed by different nodes may be uncertain in a partially synchronous network. This fork selection mechanism avoids the limitations of nondeterminism in a partially synchronous network while ensuring the security of the protocol.

\vspace{3pt}
\noindent
\emph{Security of Fork Selection. }
During the fork selection process, a malicious leader may attempt to violate the fork selection rules through various means. However, these attempts do not compromise the security of the LCE protocol. For instance, a malicious leader in an honest shard may attempt to select multiple forks (equivocation attack), choose a conflicting fork (i.e., one that is not an extension of the previously finalized fork by the LCE protocol), or refuse to produce a block. However, as mentioned, since all endorsements must be approved through intra-shard consensus, such malicious actions will be detected by honest nodes within the shard, resulting in endorsement failure and triggering a view change.
A malicious leader in an honest shard may also attempt to avoid endorsing the longest fork from its perspective. This behavior can be tolerated, as long as the leader attempts to endorse a block from any non-conflicting fork (and not multiple forks), the honest nodes in intra-shard consensus do not determine whether the fork is the longest. Therefore, this selection will pass consensus and will not compromise the protocol's security, although it may slow down the system's progress.
For corrupted shards, consensus within the shard may be used to endorse multiple forks against the rules. However, since there is at least one honest shard within each endorsement group, as described earlier, that honest shard ensures that a single fork is selected for endorsement through intra-shard consensus, thereby ensuring the overall security of the LCE protocol. A simple example is illustrated in Figure \ref{fig1}.

As shown in Figure~\ref{fig1}, without loss of generality, an endorsement group consists of $Shard_0$, $Shard_1$, and $Shard_2$.
After blocks $A_2'$ and $A_2''$ are prepared, $Shard_1$ and $Shard_2$ need to determine their endorsements in succession.
In $Shard_1$, block $B_3$ can endorse either block $A_2'$ or $A_2''$ without conflict because block $B_2$ (parent of block $B_3$) has endorsed block $A_1$ (parent of block $A_2'$ and $A_2''$).
Without loss of generality, the proposer of block $B_3$ in $Shard_1$ receives valid prepared block $A_2'$ first, so $Shard_1$ endorses block $A_2'$ in block $B_3$.
In corrupted $Shard_2$, the malicious block proposer forks the shard chain by proposing two blocks ($C_4'$ and $C_4''$), both of which can only endorse $B_3$ after $Shard_1$'s endorsement decision.

\vspace{3pt}
\noindent
\textbf{Block Finalization}.
In this phase, each shard is responsible for informing the finalized block back to its owner shard.
A prepared block will be finalized when it is endorsed by all shards in its endorsement group.
Since the order of endorsements is known, and the endorsement history is accumulated and transmitted along the endorsement order, the last shard in the endorsement order knows the original block that is fully endorsed.
In this case, the last shard must inform the original shard of the finalized original block by transmitting the accumulated endorsement history.

Once the original shard selects the block containing the latest endorsement history from the last shard in the endorsement order, the original shard's nodes would confirm the pre-execution of that finalized block and roll back (i.e., \emph{discard}) the conflicting \emph{prepared} ones.

As shown in Figure~\ref{fig1}, block $A_2'$ is prepared by $Shard_0$ (i.e., endorsed by $Shard_0$) and endorsed by block $B_3$, then block $B_3$ is endorsed by $C_4'$ or $C_4''$.
Block $A_2'$ is finalized since it is endorsed by all shards within the endorsement group.
Then, the honest nodes in $Shard_2$ transmit block $C_4'$ or $C_4''$ containing the accumulated endorsement history to $Shard_0$.
Note that the nodes in $Shard_0$ can select either block $C_4'$ or $C_4''$ since both finalize block $A_2'$.
This safe finalization stems from honest $Shard_1$'s consistent endorsement of $Shard_0$'s blocks.
Without loss of generality, $Shard_0$ selects block $C_4'$, then the nodes in $Shard_0$ confirm block $A_2'$'s pre-execution and consider the blocks from $A_2''$ to be discarded.
Finally, newly prepared blocks in $Shard_0$ can only be the child of block $A_2'$.

\begin{figure}[t]
\centerline{\includegraphics[width=0.45\textwidth]{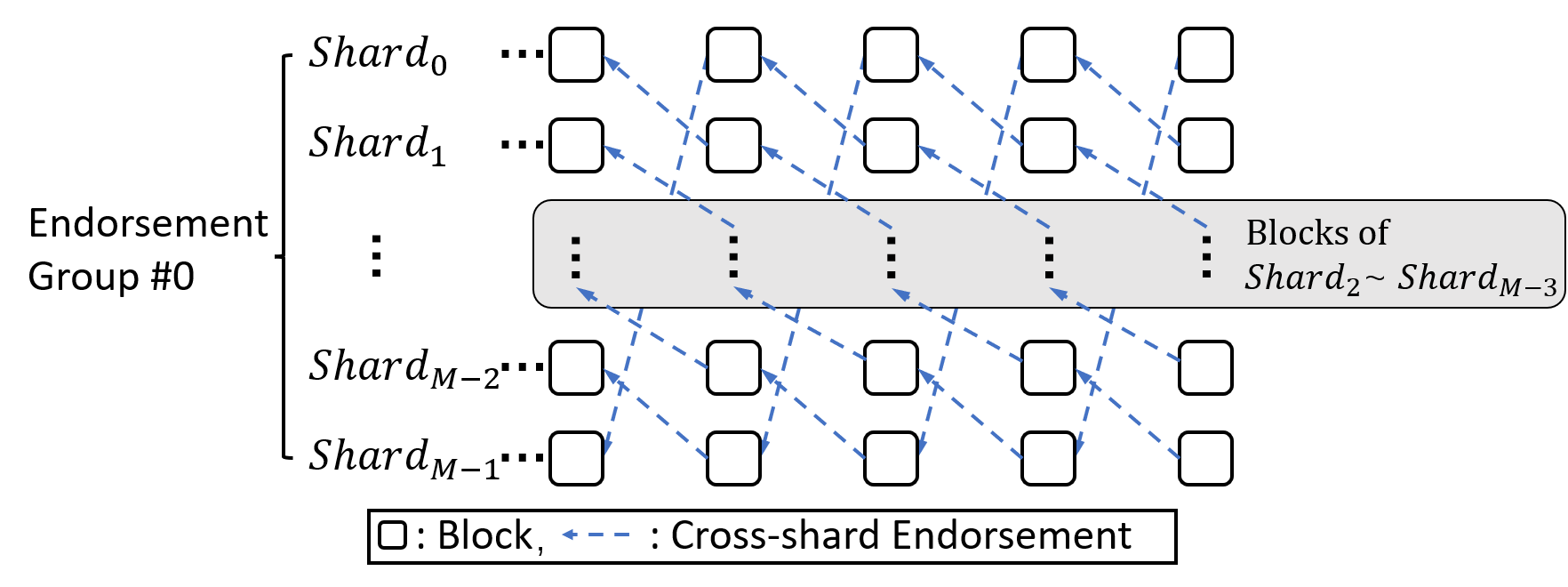}}
\vspace{-12pt}
\caption{Linked endorsements within an endorsement group of size $M$. }
\label{fis:endorsementGroup}
\end{figure} 

\subsection{Efficient Cross-shard Communication for LCE}\label{subsub:efficient_communication}
In this part, we illustrate how to reduce the overhead of cross-shard communication in \ac{LCE} by only using block headers.
Additionally, we only require block headers to be broadcast to another shard rather than the entire network.

To verify the validity of a prepared block before endorsing, the aforementioned basic design requires to store the shard state and verify each raw transaction inside that block, leading to massive overhead.
However, we hope to complete the verification before endorsement with little overhead.

To achieve this goal, our core idea is to let each shard ensure the validity of individual transactions through intra-shard consensus, while employing the LCE protocol to resolve potential forks at the shard level using only block headers. In this way, the resource-intensive task of verifying specific transactions based on shard state remains entirely within each shard. Meanwhile, the LCE protocol leverages lightweight information contained in the block headers (e.g., consensus signatures) to efficiently detect and resolve forks introduced by malicious shards.

Specifically, we fine-tune the system parameters according to the mathematical expressions in Section~\ref{epoch_security} to ensure the fraction of malicious nodes in each shard is $<2/3$, the quorum size of BFT-typed consensus, except with negligible failure probability.
In this scenario, the adversary cannot control a sufficient number of malicious nodes to approve blocks containing invalid transactions through intra-shard consensus. Specifically, when the fraction of malicious nodes is below the required quorum threshold for intra-shard consensus (i.e., 2/3 of the shard size), any attempt to reach consensus on a block containing invalid transactions—thus achieving the quorum threshold (2/3)—necessitates obtaining at least one honest node's signature. However, honest nodes verify transaction validity rigorously and consequently will never endorse such invalid transactions. Therefore, under this condition, all \emph{prepared} blocks that successfully pass intra-shard consensus must contain only valid transactions.

However, even though malicious nodes constituting less than 2/3 of the shard are unable to tamper with transactions (as above-mentioned), they can still induce forks within the shard to undermine security. Fortunately, the detection of such forks can be efficiently achieved through block headers.
Specifically, with the nodes' identity recorded in the beacon chain, each node from another shard only needs to verify whether the received block is prepared by verifying the signatures in the block header.
Moreover, we require \emph{each node} to be responsible for the cross-shard communication of the blocks it votes for.
Hence, at least one honest node will broadcast the \emph{prepared} block to the corresponding shard for subsequent endorsement and finalization.
As a result, we can \emph{reuse the block header (instead of the whole block)} with two additional fields for endorsement and finalization.
1) The \emph{latest finalized hash} indicates the hash of the latest finalized block in its shard.
2) The \emph{endorsement list} consists of hashes of up to $M-1$ blocks going back to the original shard's block, indicating the recent history of the linked endorsements.
The endorsement list is updated across each step of the linked endorsements.
For example, when a shard proposes block $B$ to endorse the latest block $A$ from the preceding shard in the endorsement group, block $B$'s endorsement list is copied from that in block $A$ with an additional hash of block $A$.

Beyond leveraging block headers for endorsement and finalization, we underscore that each shard only needs to broadcast its block headers to the next shard in the endorsement group order rather than the entire network.
The feasibility of it lies in the fact that when a shard broadcasts a block header to the subsequent shard, the block header accumulates a history of endorsements necessary to finalize a block of the subsequent shard.
Such a circular broadcasting within each endorsement group inspires the name of our system, suggesting that the cross-shard endorsements within the endorsement group are shaped like infinitely extending \emph{spirals}, as shown in Figure~\ref{fis:endorsementGroup}.
In Figure~\ref{fis:endorsementGroup}, the links between blocks within the same shard are hidden.
Specifically, if $Shard_w$ has the largest shard ID in its endorsement group of size $M$, its header would be sent to the shard having the smallest shard ID within its group,
otherwise, block headers of $Shard_w$ would be sent to $Shard_{w+1}$.

\begin{figure}[t]
\centerline{\includegraphics[width=0.36\textwidth]{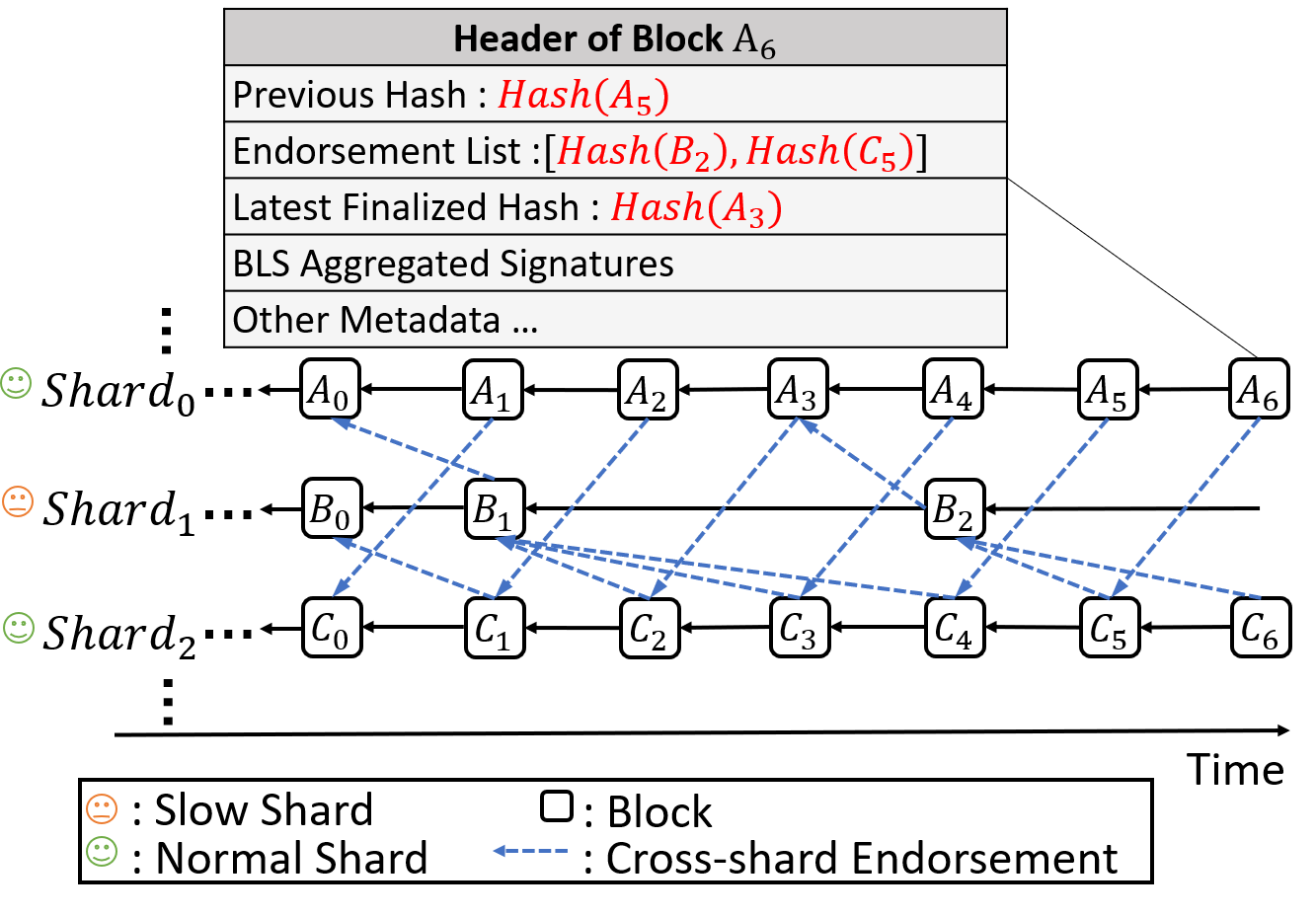}}
\vspace{-15pt}
\caption{Pipelining mechanism.}
\label{fig2}
\end{figure}

\subsection{Pipelining Mechanism}
\label{subsec:pipe}
In SpiralShard, the sequentially processed endorsement within the endorsement group may be halted due to a slow shard.
Hence, we pipeline block production so that a faster shard can package transactions at its speed.
This mechanism requires shards to produce blocks optimistically and conduct endorsement on the latest block from the preceding shard instead of each block.

During the \emph{block preparation} phase in LCE, a shard may wait until its latest prepared block has been finalized before preparing a new one.
However, in this case, once there is a slow shard leading to a slow finalization, the speed of block preparation of all the shards within the same endorsement group will be slowed down.
Hence, we require that each shard optimistically prepares blocks after non-discarded blocks instead of directly after finalized blocks.
After optimistic preparation, nodes are still required to broadcast the prepared blocks to the next shard.
In this case, even if the blocks can not be finalized in time, shards can continue to package new transactions to the blocks.

Additionally, during the \emph{cross-shard endorsement} phase in LCE, a shard may receive multiple blocks at different heights\footnote{The "height" of a block refers to the number of blocks that precede it in the same chain. For instance, the first block in the chain has a height of 0, and the next block has a height of 1, and so forth.} optimistically prepared by the preceding shard.
If a shard needs to endorse \emph{each} prepared block of the preceding shard directly,
then the finalization speed for each shard is limited by the slowest shard within the same endorsement group.
Hence, we require each shard to endorse the latest non-conflicting prepared block from the preceding shard instead of each block.
Specifically, the node verifies the validity of this endorsement decision to prevent it from conflicting with the parent block's endorsement decision (i.e., endorse on the different fork branches).

During the \emph{block finalization} phase in LCE, the block $A$ may not directly collect sufficient endorsements.
But it is also safely finalized once block $A$'s subsequent block is finalized since no other blocks conflicting with block $A$ can be sufficiently endorsed.
As a result, shards can optimistically append blocks to the chains and finalize blocks in time.

As shown in Figure~\ref{fig2},  we set the endorsement group consisting of $Shard_0$, $Shard_1$, and $Shard_2$, where $Shard_1$ is slow.
$ Shard_2$ optimistically produces blocks $C_2$, $C_3$, and $C_4$ without waiting, and all of these blocks endorse the latest non-conflicting block $B_1$ from $Shard_1$.
Then $Shard_1$ produces block $B_2$ endorsing the latest block $A_3$ from $Shard_0$.
Once block $C_5$ is produced to endorse block $B_2$, block $A_3$ is endorsed by all the shards within the endorsement group. Block $A_1$ and $A_2$ can also be considered finalized since they are parents of block $A_3$.
Honest nodes in $Shard_0$ must show the headers from block $A_2$ to block $A_6$ to prove that block $A_2$ is a parent of finalized block $A_3$ to finalize $A_2$.
As a result, once the slow shard prepares new blocks, optimistic prepared blocks from other shards can be quickly finalized to restore throughput.

%% file: Files/analysis.tex
\section{Analysis and Discussion}
\label{sec:analysis}

\subsection{Epoch Failure Probability Analysis}
\label{epoch_security}
In this section, we calculate the upper bound of the failure probability within each epoch.
In SpiralShard, we leverage cumulative hypergeometric distribution function to calculate probabilities, similar to previous works~\cite{rapidchain, pyramid, li2022jenga, cycledger, omniledger}.
Besides, previous works typically assume the failure probability of each shard is independent while adopting union bound\footnote{The union bound delivers an upper limit or a 'worst-case scenario' estimate for the occurrence of any event within a given set.} to calculate the upper bound of the system's failure probability~\cite{rapidchain, pyramid, li2022jenga, cycledger, omniledger}.
We adopt the same assumption and mathematical tools.
In SpiralShard, an endorsement group will fail in three cases:
\textbf{Case 1}: It contains no honest shard, i.e., all shards in this endorsement group have at least $1/3$ fraction of malicious nodes. 
\textbf{Case 2}: It contains honest shards, but at least one shard is corrupted by at least $2/3$ fraction of malicious nodes.
\textbf{Case 3}: It contains honest shards, but at least one shard is corrupted by at least $1/3$ fraction of Byzantine nodes.

We first calculate an endorsement group's failure probability for \textbf{\emph{case 1}}.
Here, we calculate the probability of the event that the proportion of malicious nodes in an endorsement group is at least $1/3$ as the upper bound since this is a \emph{necessary condition} for such an endorsement group not to contain any honest shard.
Note that the network size, the shard size, the endorsement group size, and the fraction of malicious nodes in the network are denoted as $N, S, G,  F$ ($F = F_A + F_B$), respectively.
Correspondingly, we denote the number of nodes per endorsement group as $M = S \cdot G$ for simplicity of formulas.
Let $X_i$ denote the random variable of the number of malicious nodes in endorsement group $i$.
The probability of endorsement group $i$ having $X_i=x$ malicious nodes is:

\vspace{-9pt}
\begin{equation}
Pr[X_i = x]
= \frac{
\binom{F\cdot N}{x} \binom{N-F\cdot N}{M-x}
}
{
\binom{N}{M}
}.
\label{group_has_x_F}
\end{equation}
\vspace{-9pt}

Based on Equation~\ref{group_has_x_F}, we can derive the upper bound of the failure probability of endorsement group $i$ for \textbf{\emph{case 1}} by letting $X_i \geq M/3$:

\vspace{-18pt}
\begin{equation}
Pr[\text{Failure}_i^{(\text{Case 1})}]
\leq  Pr[X_i \geq M/3]  = \sum_{x= \lfloor M / 3\rfloor }^{M} Pr[X_i = x].
\label{group_case_1}
\end{equation}
\vspace{-15pt}

We now calculate the failure probability of endorsement group $i$ for \textbf{\emph{case 2}}.
In endorsement group $i$, let $Y_{ij}$ denote the random variable of the number of malicious nodes in $Shard_j$. 
If given that endorsement group $i$ having $X_i=x$ malicious nodes, the probability of $Shard_j$ having $Y_{ij} = y$ malicious nodes can be expressed as:

\vspace{-12pt}
\begin{equation}
Pr[ Y_{ij} = y | X_i = x]
=  \frac{
\binom{x}{y} \binom{M-x}{S-y}
}
{
\binom{M}{S}
}.
\label{shard_has_y_F|group_has_x_F}
\end{equation}
\vspace{-9pt}

Based on Equations~\ref{group_has_x_F} and \ref{shard_has_y_F|group_has_x_F}, 
the probability that endorsement group $i$ fails due to \textbf{\emph{case 2}} caused by $Shard_j$ (i.e., $Y_{ij} \geq 2S/3$) can be expressed as:

\vspace{-12pt}
\begin{equation}
\begin{aligned}
&Pr[Y_{ij} \geq 2S/3 \cap X_i < M/3] \\
&=
\sum_{x= 1 }^{\lfloor M / 3 -1\rfloor }
\sum_{y= \lfloor 2S / 3 \rfloor }^{ S }
Pr[X_i = x] \cdot Pr[ Y_{ij} = y | X_i = x] 
. 
\end{aligned}
\label{shard_cause_case_2}
\end{equation}
\vspace{-9pt}

Based on Equation~\ref{shard_cause_case_2}, we leverage the union bound over $G$ shards within endorsement group $i$ to bound the probability of \emph{existing} a $j$ such that $Y_{ij} \geq 2S/3$, failing endorsement group $i$ for \textbf{\emph{case 2}}.

\vspace{-15pt}
\begin{equation}
Pr[\text{Failure}_i^{(\text{Case 2})}]
 \leq
G \cdot Pr[Y_{ij} \geq 2S/3 \cap X_i < M/3]
 \label{group_case_2}
.
\end{equation}
\vspace{-15pt}

We now calculate the failure probability of endorsement group $i$ for \textbf{\emph{case 3}}.
Let $Z_i$ denote the random variable of the number of Byzantine nodes in endorsement group $i$.
Similar to Equation~\ref{group_has_x_F}, the probability of endorsement group $i$ having $Z_i=z$ Byzantine nodes can be expressed as:

\vspace{-9pt}
\begin{equation}
Pr[Z_i = z]
= \frac{
\binom{F_B\cdot N}{z} \binom{N-F_B\cdot N}{M-z}
}
{
\binom{N}{M}
}.
\label{group_has_z_FB}
\end{equation}
\vspace{-9pt}

In endorsement group $i$, let $W_{ij}$ denote the random variable of the number of Byzantine nodes in $Shard_j$. 
If given that endorsement group $i$ having $Z_i=z$ Byzantine, the probability of $Shard_j$ having $W_{ij} = w$ Byzantine can be expressed as:

\vspace{-9pt}
\begin{equation}
Pr[ W_{ij} = w | Z_i = z]
=  \frac{
\binom{z}{w} \binom{M-z}{S-w}
}
{
\binom{M}{S}
}.
\label{shard_has_w_FB|group_has_z_FB}
\end{equation}
\vspace{-9pt}

Based on Equations~\ref{group_has_z_FB} and \ref{shard_has_w_FB|group_has_z_FB}, 
the probability that endorsement group $i$ fails due to \textbf{\emph{case 3}} caused by $Shard_j$ (i.e., $W_{ij} \geq S/3$) can be expressed as:

\vspace{-15pt}
\begin{equation}
\begin{aligned}
&Pr[W_{ij} \geq S/3 \cap Z_i < M/3] \\
&=
\sum_{z= 1 }^{\lfloor M / 3 -1\rfloor }
\sum_{w= \lfloor S / 3 \rfloor }^{ S }
Pr[Z_i = z] \cdot Pr[ W_{ij} = w| Z_i = z] 
. 
\end{aligned}
\label{shard_cause_case_3}
\end{equation}
\vspace{-9pt}

Based on Equation~\ref{shard_cause_case_3}, we leverage union bound over $G$ shards within endorsement group $i$ to bound the probability of \emph{existing} a $j$ such that $W_{ij} \geq S/3$, failing endorsement group $i$ for \textbf{\emph{case 3}}, similar to Equation~\ref{group_case_2}.

\vspace{-15pt}
\begin{equation}
Pr[\text{Failure}_i^{(\text{Case 3})}]
 \leq
 G \cdot Pr[W_{ij} \geq S/3 \cap Z_i < M/3]
 \label{group_case_3}
.
\end{equation}
\vspace{-15pt}

Although \textbf{cases 2} and \textbf{3} are not mutually exclusive, we use union bound over three cases to bound the failure probability of endorsement group $i$ for simplicity as follows:

\vspace{-12pt}
\begin{equation}
\begin{aligned}
Pr[\text{Failure}_i] \leq 
&Pr[\text{Failure}_i^{(\text{Case 1})}]   + Pr[\text{Failure}_i^{(\text{Case 2})}] \\
&\quad + Pr[\text{Failure}_i^{(\text{Case 3})}].
\end{aligned}
\end{equation}
\vspace{-9pt}

Accordingly,  we leverage the union bound over $N/M$ groups to calculate the probability of \emph{existing} an $i$ such that the system fails due to endorsement group $i$, as follows:

\vspace{-12pt}
\begin{equation}
 \begin{aligned}
&Pr[\text{System Failure} ] \leq N/M \cdot Pr[\text{Failure}_i] 
\label{system_fail}
\end{aligned}
\end{equation}
\vspace{-15pt}

\noindent
\textbf{Ensuring Negligible Failure Probability.} 
SpiralShard must ensure a negligible failure probability within each epoch to maintain security, similar to most blockchain sharding works~\cite{omniledger, rapidchain, cycledger, pyramid, li2022jenga}. 
Based on Equations~\ref{group_has_x_F},~\ref{shard_cause_case_2}, and \ref{shard_cause_case_3}, we need to adjust the shard size $S$ and endorsement group size $G$ to make sure there is a small $\varepsilon$ existing so that:

\vspace{-9pt}
\begin{equation}
\begin{aligned}
\varepsilon &\geq N/M \cdot (Pr[X_i \geq M/3] \\
&\quad + G \cdot Pr[Y_{ij} \geq 2S/3 \cap X_i < M/3] \\
&\quad + G \cdot Pr[W_{ij} \geq S/3 \cap Z_i < M/3]).
\label{final_fail_pro}
\end{aligned}
\end{equation}
\vspace{-9pt}

The specific values of the probability can be found in Table~\ref{table:parameter}.


\subsection{Protocol Security Analysis}
\label{protocol_security}

Based on negligible epoch failure probability, we now analyze the safety and liveness of our \ac{LCE} protocol.

 
\begin{theorem}
\label{Theorem2}
The \ac{LCE} protocol provides safety if each corrupted shard has less than $2/3$ malicious nodes (i.e., less than $1/3$ fraction of Byzantine nodes plus no more than $1/3$ fraction of a-b-c nodes) and at least one honest shard exists in each endorsement group (i.e., each endorsement group has less than $1/3$ of malicious nodes) within the same epoch.
\end{theorem}

\begin{proof}
We use proof by contradiction to demonstrate this theorem.
Without loss of generality, we demonstrate the safety of a specific finalized block $B_a$.
Since block $B_a$ is finalized, it has been verified by at least one honest shard and has received endorsements from all shards in its endorsement group within the epoch.
Assuming another block, $B_b$, which conflicts with $B_a$, is finalized in the same epoch.
This assumption is based on two distinct scenarios.
\emph{First}, all shards in the endorsement group simultaneously endorse these two blocks via intra-shard consensus.
This means, in this case, there is no honest shard in the endorsement group (i.e., the proportion of malicious nodes in the endorsement group is at least $1/3$).
However, this contradicts the theorem that at least one honest shard exists in each endorsement group.
\emph{Second}, a corrupted shard manipulates the other shards' endorsement decisions by setting an invalid endorsement list in its block header during intra-shard consensus.
This requires the number of malicious nodes within this corrupted shard to be larger than the quorum size of BFT-typed consensus.
However, it contradicts the theorem that the fraction of malicious nodes within each shard is less than $2/3$.
In both scenarios, the theorem is violated.
Therefore, if the conditions of the theorem are met, the safety is preserved without losing generality.

Notably, the presence of malicious nodes, which exceed 1/3 but remain below 2/3 in a shard, can potentially introduce nondeterminism during the view change process within corrupted shards. For instance, due to the presence of a malicious leader or a large number of malicious nodes, these nodes may deliberately manipulate information, causing some honest nodes to believe that the view change has been completed while others do not perceive a need for it. This may lead to inconsistent states among honest nodes and the persistent formation of forks.
In SpiralShard, we tolerate such nondeterminism within corrupted shards, as forks resulting from nondeterminism are ultimately resolved by the LCE protocol, which selects a single branch to finalize (Section \ref{subsec:basicOCE}). Once a fork in a corrupted shard is finalized through the LCE protocol, honest nodes within the shard that had previously chosen a different branch (due to nondeterminism) will roll back their state to the latest finalized state determined by the LCE protocol. They will then proceed with state updates based on the branch selected by the current LCE protocol. Overall, any nondeterminism arising from intra-shard consensus or view change processes within corrupted shards can eventually be resolved by the LCE protocol, which finalizes only one of the forks.

\end{proof}


\begin{theorem}
\label{Theorem:liveness}
The \ac{LCE} protocol provides liveness if each corrupted shard has less than $1/3$ fraction of Byzantine nodes plus no more than $1/3$ fraction of a-b-c nodes and at least one honest shard exists in each endorsement group within the same epoch.
\end{theorem}

\begin{proof}
We first prove that the intra-shard BFT-typed consensus can maintain liveness, and then, based on this, prove that the LCE protocol can maintain liveness.

First, for the intra-shard consensus. 
In accordance with the definition of a-b-c nodes, elaborated in Section~\ref{subsec:threat_model} and supported by the existing work~\cite{flexible}, these nodes preserve liveness.
Indeed, as demonstrated in Theorem~\ref{Theorem2}, safety is provided. 
Consequently, once safety is established, the a-b-c nodes would be treated as honest and would not prevent intra-shard consensus (also view change) process from being proceeded,
aligning with the principles of a-b-c nodes in the existing work~\cite{flexible}.
Since no more than 1/3 of the nodes in each shard are a-b-c nodes, and these nodes do not attack liveness, the proportion of nodes attacking liveness in each shard remains below 1/3, which is within the fault tolerance threshold of BFT-typed consensus. 
Therefore, within each shard, both the consensus process and the view change process can maintain liveness.

Second, for the LCE protocol. 
As mentioned above, the consensus and view change processes within each shard maintain liveness. This means that each shard can continuously endorse other shards through its consensus process. Furthermore, based on Theorem \ref{Theorem2} and Section \ref{subsec:basicOCE}, an honest shard will always select a single fork for endorsement through intra-shard consensus. Given these two points, the LCE protocol requires that shards sequentially endorse the preceding shard in each endorsement group.
In this case, each shard only needs to endorse one previous shard directly and can implicitly endorse the endorsement decisions of the preceding shard. 
When a shard endorses, it is aware of which fork its preceding shard has endorsed, allowing honest shards to make non-conflicting endorsements. 
Therefore, in the LCE protocol, the sequential endorsements between shards prevent situations where different honest shards endorse different forks, leading to insufficient endorsements for any particular fork. 
Consequently, through the LCE protocol, honest shards within each endorsement group sequentially endorse the same fork. 
This ensures that, within the same endorsement group, there will always be one fork that collects endorsements from all shards, thereby guaranteeing the liveness of the LCE protocol.

\end{proof}

\subsection{Corner Cases Discussion}

Apart from typical attacks (i.e., transaction manipulation, double-spending, silent attack), DDoS attacks may be launched by Byzantine nodes or attackers outside the blockchain network to break liveness.
For Byzantine nodes within the system, for example, Byzantine leaders may propose too many forking blocks to consume other nodes' network bandwidth, compromising the liveness.
Fortunately, many robust BFT protocols today (such as Tendermint \cite{buchman2016tendermint} or Harmony's FBFT \cite{Harmony}) typically incorporate multiple mechanisms to resist or mitigate such attacks. For instance, nodes can apply bandwidth or rate limits to incoming blocks, or use 'gossip by demand,' allowing nodes to request blocks from the leader only when needed, rather than having the leader proactively broadcast all blocks. Additionally, each node can quickly detect whether the leader has sent multiple blocks through signature verification. Upon detecting equivocation, nodes can discard all subsequent messages from that leader.
Regarding external attackers, such as those attempting to launch attacks by submitting numerous spam transactions or sending a large number of invalid data packets to blockchain nodes, there are several classic defense measures. Examples include employing transaction fees to increase attack costs, applying traffic control to RPC nodes, or using DDoS protection tools (e.g., Cloudflare, AWS Shield) to protect the external IP addresses of nodes. 

Censorship attacks represent another possible threat, where a malicious leader selectively chooses not to include specific transactions in the blocks \cite{mev}. Such attacks can be mitigated through various additional measures. For instance, privacy-enhancing technologies \cite{javed2021petchain} can be used to conceal the content of transactions, making them less susceptible to censorship. Alternatively, leaderless BFT-type consensus protocols can be employed. 
For instance, Red Belly~\cite{redBelly} involves merging micro-blocks proposed by multiple nodes into a unified block, thus ensuring the integrity of transactions. 
Additionally, this consensus protocol offers defense against eclipse attacks targeting specific leaders. 

{\color{red}
In summary, most of the aforementioned attacks, along with other potential threats, can typically be mitigated through well-established standard defense mechanisms. As these mechanisms are specifically designed to counter their respective attack vectors, they help ensure that system performance remains largely unaffected. Importantly, these lines of research are orthogonal to the core design of SpiralShard. Nonetheless, SpiralShard can be readily integrated with such existing solutions to enhance its resilience against atypical attacks while maintaining system performance.
}

\subsection{Cross-shard Smart Contract}
\label{subsec:cx_contract}

For cross-shard smart contract processing, we make slight modifications inspired by the Two-Phase Commit (2PC) protocol~\cite{sgxSharding}.
Specifically, a shard packages a \texttt{BeginTX} into a block when a client requests to call its smart contract.
Once this block is finalized, the shard sends the \texttt{BeginTX} with finality proof (Merkle tree path and block header) to relevant shards to lock states.
Upon verifying the proof, corresponding shards package \texttt{PrepareOK} messages to lock states (or \texttt{PrepareNotOK} messages in cases where locking isn't feasible).
The finalized \texttt{PrepareOK/PrepareNotOK} messages are sent back to the primary shard with finality proof.
This primary shard will package a \texttt{CommitTX} to execute the original smart contract if all the required states are locked (or \texttt{AbortTX} messages otherwise).
Once the \texttt{CommitTX/AbortTX} is finalized, it is broadcast back to the corresponding shards with the proof of finality to execute the commit or abort operation.
It is worth noting that every message (\texttt{BeginTX}, \texttt{PrepareOK}, \texttt{PrepareNotOK}, \texttt{CommitTX}, and \texttt{AbortTX}) is transmitted with an accompanying finality proof to preserve atomicity. 

One limitation of SpiralShard is that it may increase transaction confirmation latency, as (cross-shard) transactions need to wait for finalization by the LCE protocol before they can be confirmed (Section~\ref{subsec:other}). This is especially true for cross-shard smart contract transactions, which may involve multiple rounds of finalization delays. However, as the main goal of this work is to improve concurrency, reducing transaction latency is left as a consideration for future research.


%% file: Files/evaluation.tex
\section{Evaluation}
\label{sec:evaluation}

\subsection{Implementation and Experimental Setup}
We implement a prototype of SpiralShard in Golang based on Harmony~\cite{Harmony}, a well-known permissionless blockchain sharding project once with top 50 market cap in cryptocurrency.
For a balanced evaluation, we adopt Harmony as the primary baseline system (hereafter \textit{Baseline}) and benchmark it against SpiralShard. Using Harmony isolates the impact of our design decisions and highlights the performance benefits attributable to SpiralShard's core mechanisms. 
In fact, SpiralShard's mechanisms can be seamlessly incorporated into many existing sharding frameworks \cite{elastico, omniledger, rapidchain, sgxSharding, pyramid, cycledger}. 
In addition, we implement a prototype of GearBox \cite{gearbox} (CCS '22)--a representative state-of-the-art sharding protocol that explicitly tolerates corrupted shards--to provide a contemporary point of comparison in our performance study.


We deploy a network with up to $4,200$ nodes on $21$ Amazon EC2 instances, each having a 96-core CPU and up to 20 Gbps network bandwidth.
Each node has 50 Mbps individual network bandwidth and 100 ms latency for each message.
We set the transaction size as $512$ bytes and let each block contain up to $4,096$ transactions.
The system's \emph{total} fraction of malicious nodes is set to 1/4, including 1/8 Byzantine nodes and 1/8 a-b-c nodes.
{\color{red}
In most of the following experiments, by default, these malicious nodes will try to generate a fork constantly.
Specifically, a malicious leader can attempt to fork the shard’s chain whenever it proposes a block.
As a result, with a 1/4 malicious node ratio, this yields a fork attempt roughly 25\% of the time.
Moreover, forking is not the only attack scenario evaluated.
For example, Section \ref{subsec:slow_shard} presents a silent attack scenario, and Section \ref{subsec:fluctuation} evaluates performance under adversarial network latency (attackers trying to delay the network).
}

In our evaluation, the transactions are generated based on the historical data of Ethereum~\cite{xblock}.
To be more realistic, the proportion of cross-shard transactions in the experiments increases with the number of shards.
This means that, given the same network scale, SpiralShard can configure a greater number of smaller shards, resulting in a higher proportion of cross-shard transactions compared to the baseline. We choose to conduct experiments under this realistic cross-shard transaction ratio (rather than controlling the cross-shard transaction ratio to be identical between SpiralShard and the baseline) mainly to facilitate a more fair and authentic performance comparison.

\subsection{Parameter Settings}
\label{parameter}

The parameter settings are shown in Table \ref{table:parameter}.
In SpiralShard, the endorsement group and shard sizes are the primary system parameters that are configured by the developers.
We require the calculated upper bound to be less than $2^{-17}\approx7.6\times10^{-6}$, meaning that the system fails once in 359 years, the same as the existing systems~\cite{pyramid, li2022jenga}.
In the baseline (Harmony), we set the shard size under the negligible failure probability of the system according to the classical hypergeometric distribution function (e.g., the equations in~\cite{rapidchain}), as the baseline does not have our LCE design.
In GearBox, we do not provide a failure probability since the framework itself lacks a methodology for calculating system-wide failure rates. 
Instead, we strictly follow the shard-scaling approach outlined in the GearBox paper by gradually increasing the shard size from small to large. After four such increases (as specified in their paper \cite{gearbox}), we compute the number and size of shards that can be successfully configured in their system.
Moreover, to ensure evaluation fairness, we configure each node in GearBox to participate in one shard, the same as we do. 

\begin{figure}[t]
\centering{}
\subfloat[Total TPS]{\includegraphics[scale=0.27]{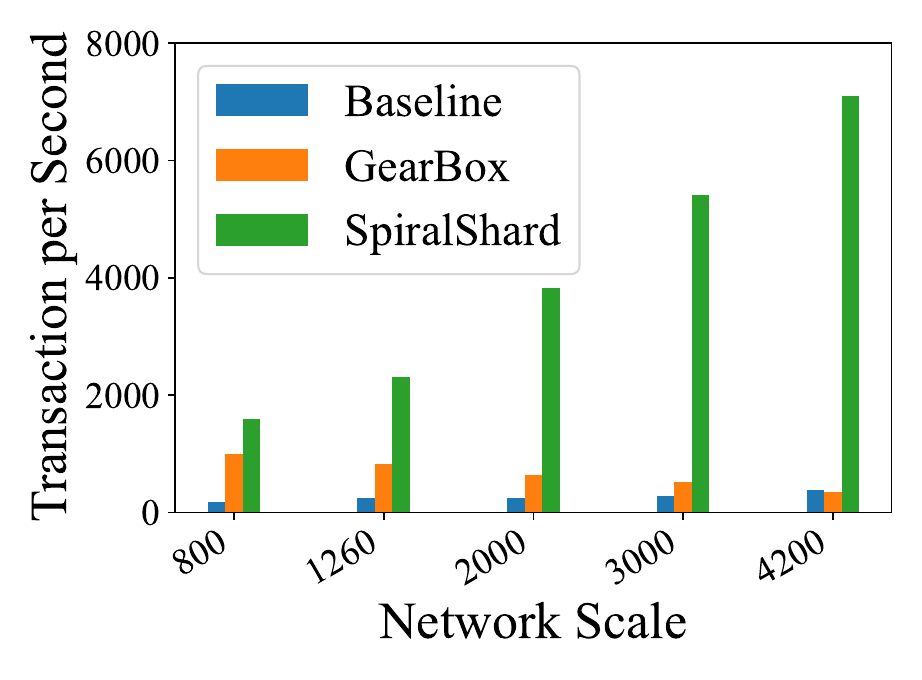}\label{subfig:throughput}}
\subfloat[Confirmation latency]{\includegraphics[scale=0.27]{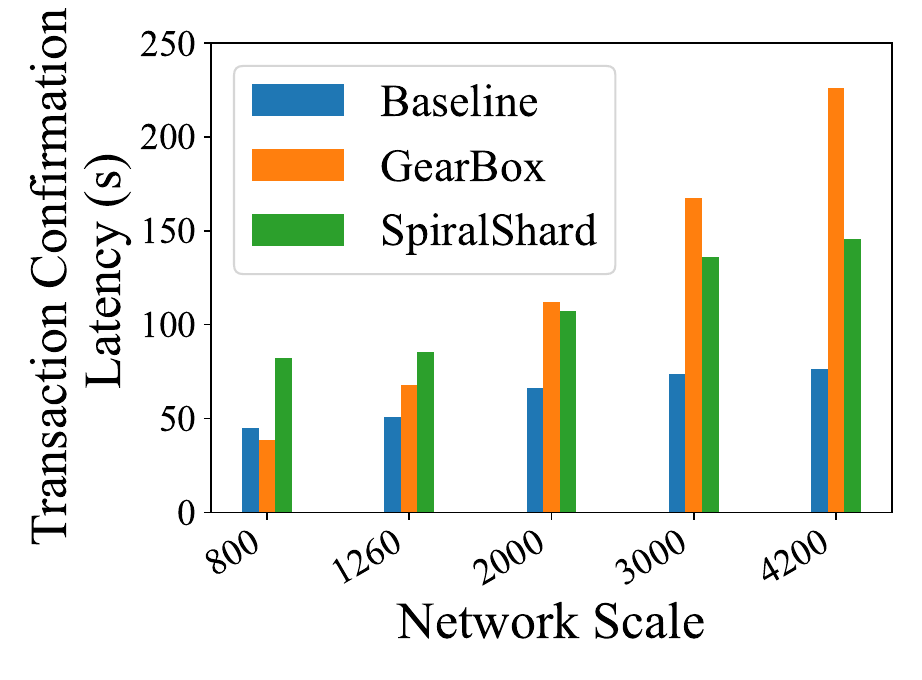}\label{fig:latency}}
\centering
\vspace{-6pt}
\caption{Performance comparison under various network sizes.}
\end{figure}

By default, we choose the shard size of around 100 for SpiralShard, which is practical and used by some previous works~\cite{rapidchain, omniledger, pyramid, sgxSharding}.
Indeed, SpiralShard can be configured with even more amount of smaller shards while maintaining the same level of security guarantees.
According to Table~\ref{table:parameter}, SpiralShard significantly reduces the shard size under the negligible failure probability, increasing the number of shards under the same network size.
It is worth noting that GearBox fails to successfully configure a large number of secure shards. This is primarily because, after the shard-scaling process, many shards still contain a fraction of malicious nodes exceeding their fault tolerance threshold, resulting in widespread shard configuration failures.

\begin{table}[h]
\caption{Parameter settings in Harmony and SpiralShard.}
\vspace{-5pt}
\centering{}%
\begin{tabular}{c|c|c|c|c|c}
\hline
\hline
Network Size & 800 & 1260 & 2000 & 3000 & 4200 \tabularnewline
\hline 
\hline
\multicolumn{6}{c}{Baseline (Harmony)}\tabularnewline
\hline
Shard Size & 400 & 420 & 500 & 600 & 600 \tabularnewline
\hline 
\# of Shards & 2  & 3 & 4 & 5 & 7 \tabularnewline
\hline 
Failure Probability ($\cdot10^{-6}$) & 0.9 & 3.9 & 4.3 & 0.8& 3.1 \tabularnewline 
\hline
\multicolumn{6}{c}{GearBox}\tabularnewline
\hline
Shard Size & 69 & 76 & 77 & 82 & 82 \tabularnewline
\hline 
\# of Shards & 5  & 8 & 11 & 14 & 14 \tabularnewline
\hline
\hline
\multicolumn{6}{c}{SpiralShard}\tabularnewline
\hline
Shard Size &100 & 105 & 100 & 100 & 100 \tabularnewline
\hline 
\# of Shards & 8 & 12 & 20 & 30 & 42 \tabularnewline
\hline 
Size of Endorsement Group &4  & 4 & 5 & 6 & 6\tabularnewline
\hline 
\# of Endorsement Group  &2 & 3 & 4 & 5 & 7 \tabularnewline
\hline 
Failure Probability($\cdot10^{-6}$)  & 0.1 & 6.5 & 4.9 & 2.1 & 5.3\tabularnewline 
\hline
\hline
\end{tabular}
\label{table:parameter}
\end{table}


\subsection{Throughput}

To measure transaction concurrency, we choose throughput as the metric.
We compare the average throughput in terms of transactions per second (TPS) at different network sizes.
As shown in Figure \ref{subfig:throughput}, 
compared to the Baseline, SpiralShard achieves up to 19.5$\times$ throughput gain at a network size of 3,000.
Compared to GearBox, SpiralShard achieves up to an 20.3$\times$ improvement in throughput (at a network scale of 4,200). Moreover, GearBox does not exhibit significant performance gains as the network size increases. This is primarily due to its reliance on network-wide consensus, which incurs substantial overhead and severely limits its scalability.
In SpiralShard, the LCE protocol is lightweight and operates without relying on network-wide consensus, resulting in an almost linear increase in TPS. This demonstrates the excellent scalability of SpiralShard.
The main reasons behind the high concurrency gain are three-fold.
First, the larger number of shards in SpiralShard brings higher transaction concurrency. 
Second, SpiralShard achieves reduced shard size, resulting in a faster intra-shard consensus.
Third, the pipelining mechanism allows each shard to continuously produce blocks without being halted by the slower shards.


Notably, SpiralShard achieves considerable performance improvements even with a higher cross-shard transaction ratio compared to the baseline. This indicates that the performance advantages of SpiralShard's unique system design outweigh the negative impacts of increased cross-shard transaction ratios. It also clearly suggests that, if the cross-shard transaction ratio between SpiralShard and the baseline are artificially controlled to be equal (possibly by employing some mechanisms to reduce the cross-shard transaction ratio \cite{LBCHAIN, brokerchain}), SpiralShard could achieve even better performance.

\subsection{Latency}
\label{sec:latency}
We compare the average transaction confirmation latency of SpiralShard and the baseline under different network sizes.
Transaction confirmation latency here is the duration from when a transaction starts to be processed until the transaction is confirmed, the same as Monoxide \cite{monoxide}.
As shown in Figure \ref{fig:latency}, compared to the Baseline, SpiralShard has longer latency (up to 1.9$\times$ at a network size of 4,200).
The main reason is that, to ensure the security of the transaction, SpiralShard requires that the transaction not be finalized until it has been endorsed by the endorsement group. 
This somewhat slows down the time it takes for a transaction to be confirmed.
However, on the other hand, although SpiralShard requires each block to be sequentially endorsed by other shards, the significantly reduced shard size also speeds up intra-shard consensus, shortening the latency for each step of the LCE, resulting in an acceptable overall latency.
Compared to GearBox, SpiralShard demonstrates lower latency when the network scales beyond 2,000 nodes (e.g., 64.4\% at a network size of 4,200). 
This is still mainly because GearBox requires network-wide consensus for block finalization.
Differently, SpiralShard only requires LCE within each endorsement group and limits the broadcasting of raw transactions within each shard.

\begin{figure}[t] 
\centering{}
\subfloat[Total TPS]{\includegraphics[scale=0.27]{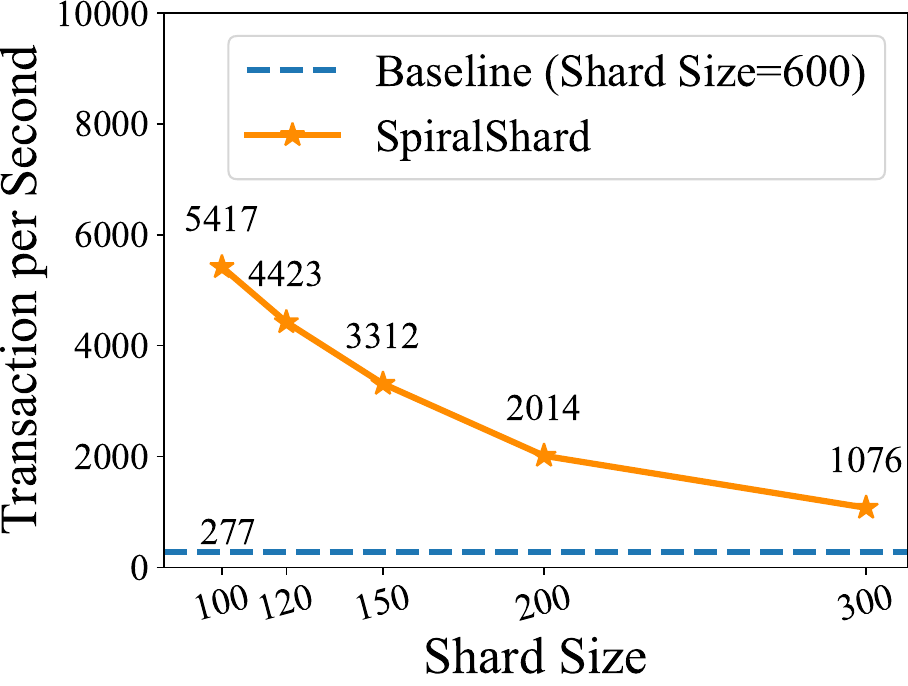}\label{subfig:tpswhen3000node}}
\hspace{5pt}
\subfloat[Confirmation latency]{\includegraphics[scale=0.27]{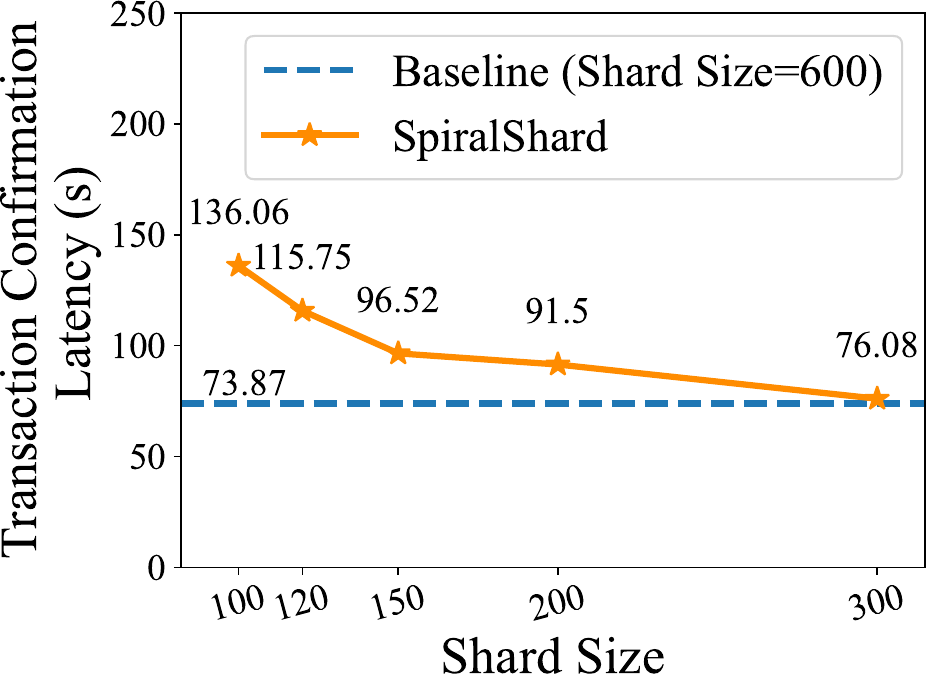}\label{subfig:latencywhen3000node}}
\vspace{-6pt}
\caption{Performance under various shard sizes.} 
\label{fig:performance3000node}
\end{figure}

\subsection{Performance under Various Shard Sizes}
\label{subsec:Eva_various_shard_size}

In SpiralShard, we can configure multiple combinations of shard and endorsement group sizes flexibility under negligible failure probabilities, which is impossible in traditional blockchain sharding systems. 
In this experiment, we fix the network size to 3,000 nodes and adjust the shard and endorsement group sizes to evaluate the system performance.
The corresponding parameter settings are shown in Table \ref{table:parameterunder3000node}.

\begin{table}[h] 
\caption{Parameters settings in SpiralShard under various shard sizes.}
\vspace{-5pt}
\centering{}%
\begin{tabular}{c|c|c|c|c|c}
\hline
\hline 
Shard Size & 100 & 120 & 150 & 200 & 300 \tabularnewline
\hline 
\# of Shards  & 30 & 25 & 20 & 15 & 10 \tabularnewline
\hline 
Size of Endorsement Group   & 6 & 5 & 4 & 3 & 2\tabularnewline
\hline 
\hline
\end{tabular}
\label{table:parameterunder3000node}
\end{table}

As shown in Figure \ref{subfig:tpswhen3000node}, the overall throughput of SpiralShard decreases as the shard size increases. 
This is because the increased shard size slows the intra-shard consensus and reduces the number of shards, threatening the system's concurrency.
However, SpiralShard still has better throughput. 
For example, even if the shard size is 300, the throughput of SpiralShard is still 3.8 $\times$ higher than that of the baseline. 
For the baseline protocol, the shard size cannot be changed arbitrarily to meet the system's success probability at the same network size, so its throughput remains the same.

The transaction confirmation latency is shown in Figure \ref{subfig:latencywhen3000node}. 
Somewhat counterintuitively, as the shard size increases, the transaction confirmation latency in SpiralShard also decreases. 
The main reason is that although the increased shard size slows down the shard consensus, it decreases the size of each endorsement group.
As a result, a block only needs to be endorsed by fewer shards, leading to lower transaction confirmation latency.

Based on the above experimental results, SpiralShard provides developers with flexible sharding configurations to suit different application scenarios. For instance, some scenarios involve large volumes of transaction data but have low demands for immediate processing. Typical examples include cross-border payments and settlements, which can tolerate delays ranging from a few minutes to hours but require high throughput to handle large-scale transaction demands, or use cases such as supply chain tracking and IoT device logging. For these scenarios, developers can configure a larger number of smaller shards (e.g., the left area in Figure \ref{subfig:tpswhen3000node} and \ref{subfig:latencywhen3000node}) to achieve higher throughput.
On the other hand, for scenarios that impact user experience, such as blockchain-based login verification, access control, or on-chain voting and governance—applications that are relatively sensitive to latency—developers can configure larger shards (e.g., the right area in Figure \ref{subfig:tpswhen3000node} and \ref{subfig:latencywhen3000node}), trading some throughput for reduced latency.
More importantly, given the flexibility of SpiralShard's sharding configurations, developers could even design mechanisms to dynamically adjust the sharding configuration within the system, thereby adapting to changing application demands in real time. We consider this direction as a potential avenue for future research.

\subsection{Overhead of LCE Protocol}

\begin{figure}[t]
\centerline{\includegraphics[width=0.24 \textwidth]{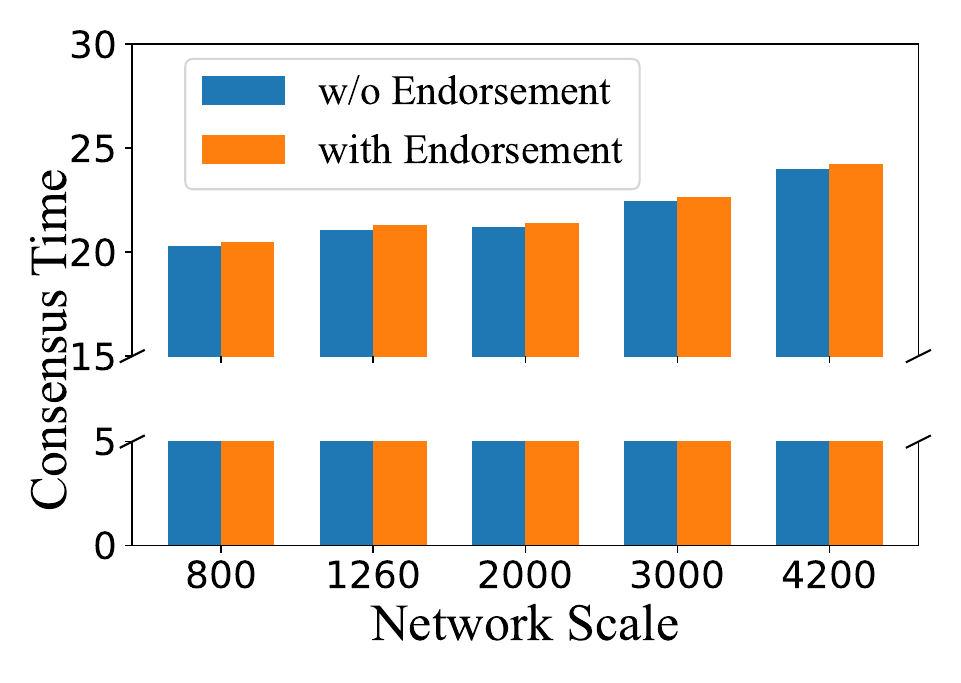}}
\vspace{-15pt}
\caption{Consensus time comparsion.}
\label{fig:woCXEndorse}
\end{figure}

We now evaluate the impact of LCE protocol's communication overhead on consensus performance. Consensus time (or speed) serves as a fundamental metric for assessing system performance, as it directly influences both throughput and transaction latency.
Specifically, we isolate the cost introduced by the LCE's broadcasting and measure the time consumed by intra-shard consensus under normal conditions. As illustrated in Figure~\ref{fig:woCXEndorse}, the communication overhead caused by cross-shard endorsement is negligible ($\approx1\%$). 
This is because, propagating cross-shard transactions already requires each shard to broadcast block headers to other shards. 
In SpiralShard, cross-shard endorsement simply reuses these block headers, adding only a few integer-sized data fields (see Figure~\ref{fig1}).

\subsection{Performance During an Epoch Transition}
\begin{figure}[t]
\centering{}
\subfloat[Total TPS]{\includegraphics[scale=0.27]{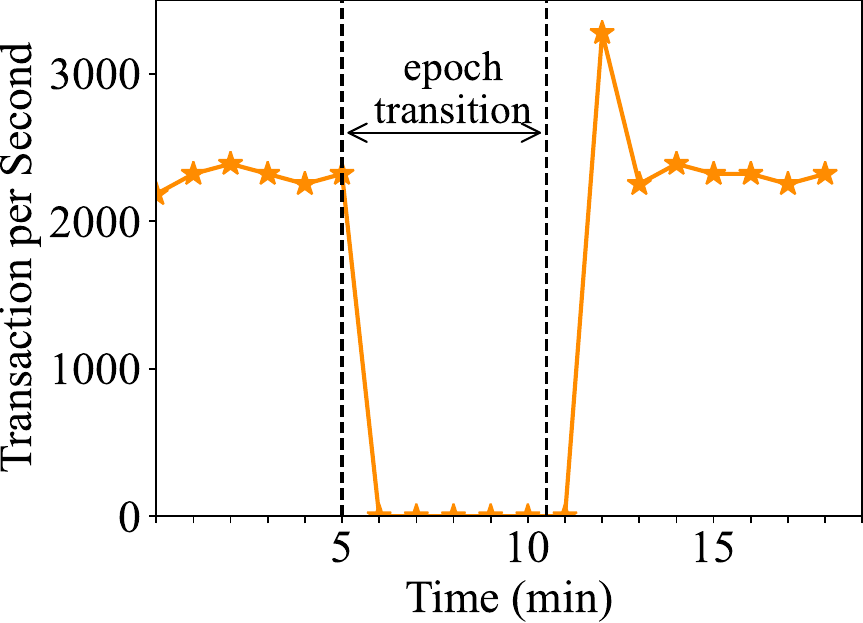}\label{subfig:tpsAcrossEpoch}}
\hspace{5pt}
\subfloat[Confirmation latency]{\includegraphics[scale=0.27]{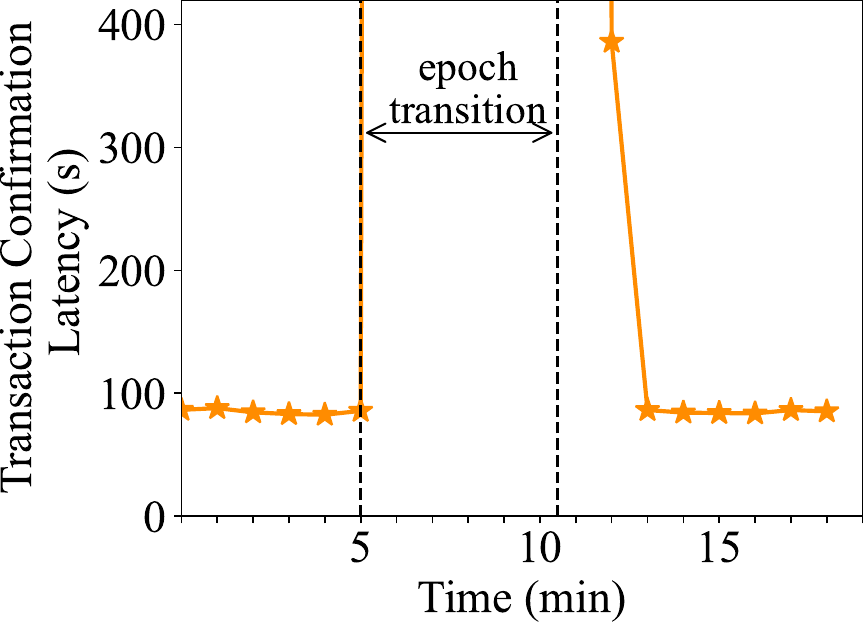}\label{subfig:latencyAcrossEpoch}}
\centering
\vspace{-6pt}
\caption{Performance during an epoch transition.}
\end{figure}

In this section, we evaluate the performance during an epoch transition (under 1,260 nodes). 
Specifically, as the system consistently finalizes blocks, we trigger an epoch transition at $T=5$. 
As shown in Figure~\ref{subfig:tpsAcrossEpoch}, there is a sharp drop in the throughput following the onset of the epoch transition. 
This is primarily due to state synchronization conducted by nodes before joining new shards.
The subsequent peak in throughput stems from the new epoch's finalizing prepared blocks from the previous epoch. 
As observed in Figure~\ref{subfig:latencyAcrossEpoch}, there is a peak in latency after the conclusion of the epoch transition. 
The main reason is that blocks generated toward the end of an epoch necessitate sequential endorsement from scratch in the new epoch.
These findings underscore that system performance experiences an impact during epoch transitions. 
However, since epoch transitions occur only once daily, aligning with existing works~\cite{rapidchain, omniledger, pyramid, li2022jenga}, the performance remains unaffected most of the time.

\subsection{Performance with Slow Shards}
\label{subsec:slow_shard}

\begin{figure}[t]
\centerline{\includegraphics[width=0.30 \textwidth]{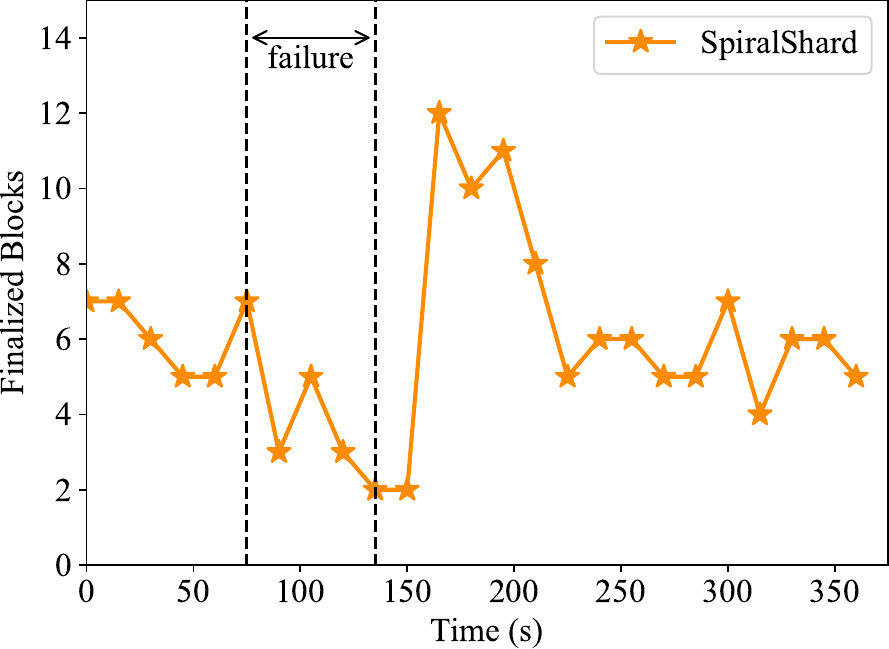}}
\vspace{-9pt}
\caption{Average finalized block (over 15s intervals) over time when network partition for 60s starting at time 75s.}
\label{fig:pipeLineDuration}
\end{figure}

{\color{red}
In this experiment, we evaluate the performance of SpiralShard in the presence of slow shards.
Specifically, we partition the network inside a shard for a while to reflect the situation such as temporary consensus failures caused by potential problems like \emph{silent attacks} launched by malicious nodes.
}

As illustrated in Figure \ref{fig:pipeLineDuration}, we start the network partition in a shard at 75s and continue for 60s. 
It is shown that the number of finalized blocks gradually decreases during this period. 
The reason is that since our LCE protocol requires sequential endorsement between shards within the endorsement group, a slow shard will halt the LCE's processing within the endorsement group. 
This results in a decreasing number of blocks that can be finalized.
Additionally, to fully reflect the impact of slow shard within the endorsement group, we have configured only one endorsement group.
After the network partition period, the consensus process for that slow shard resumes, and the number of finalized blocks increases rapidly. 
This is because, in the LCE protocol, when the newest block collects enough endorsements, all of it and its parent blocks are finalized. 
Therefore, those blocks not finalized during the network partition period will be finalized together rapidly after the consensus is restored.
{\color{red}
Noted that SpiralShard’s safety and liveness remain intact. 
In the silent attack case, no conflicting blocks are ever finalized, preserving safety.
In addition, once the affected shard recovers, all pending blocks are finalized in sequence, showing that liveness is only temporarily hindered but not permanently lost. 
}

\subsection{Performance under Network Fluctuation}
\label{subsec:fluctuation}

{\color{red}
In this section, we evaluate the performance under adversarial network fluctuation to assess the robustness (under 800 nodes with 1 endorsement group).
This evaluation can reflect the situation such as attacters \emph{delaying the network}.
}
Specifically, we use different probability distributions to model the additional delay to the propagation of each message.
These distributions are parameterized as follows: a uniform distribution with a range of 100 to 300 ms, a normal distribution with a mean (\(\mu\)) of 200 ms and a standard deviation (\(\sigma\)) of 50 ms, and a log-normal distribution where the logarithm of the delay follows a normal distribution with \(\mu = 5.268\) and \(\sigma = 0.246\). In contrast, the default case serves as a baseline with no additional propagation delay applied.

\begin{figure}[t]
\centering{}
\subfloat[Total TPS]{\includegraphics[scale=0.27]{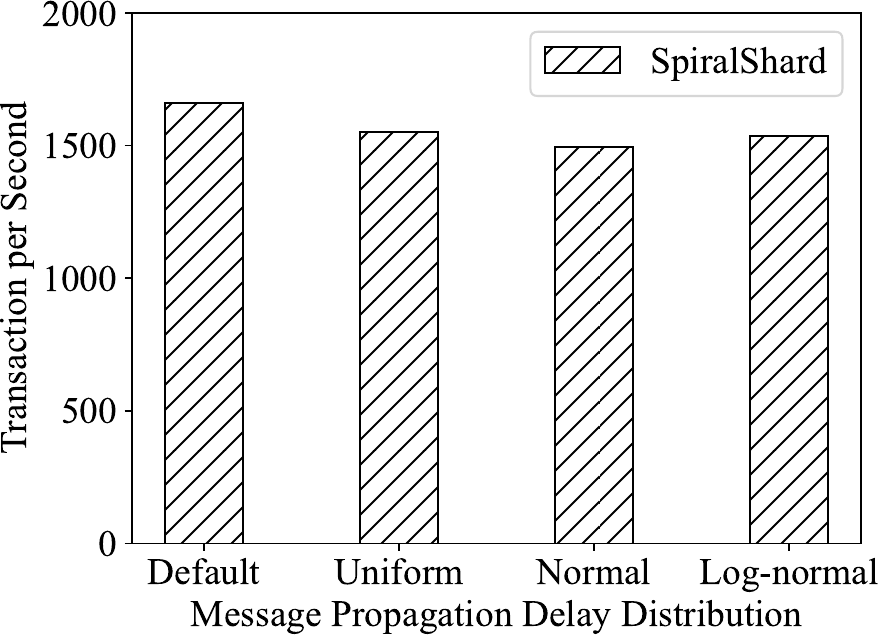}\label{subfig:throughputWithDelay}}
\hspace{5pt}
\subfloat[Confirmation latency]{\includegraphics[scale=0.27]{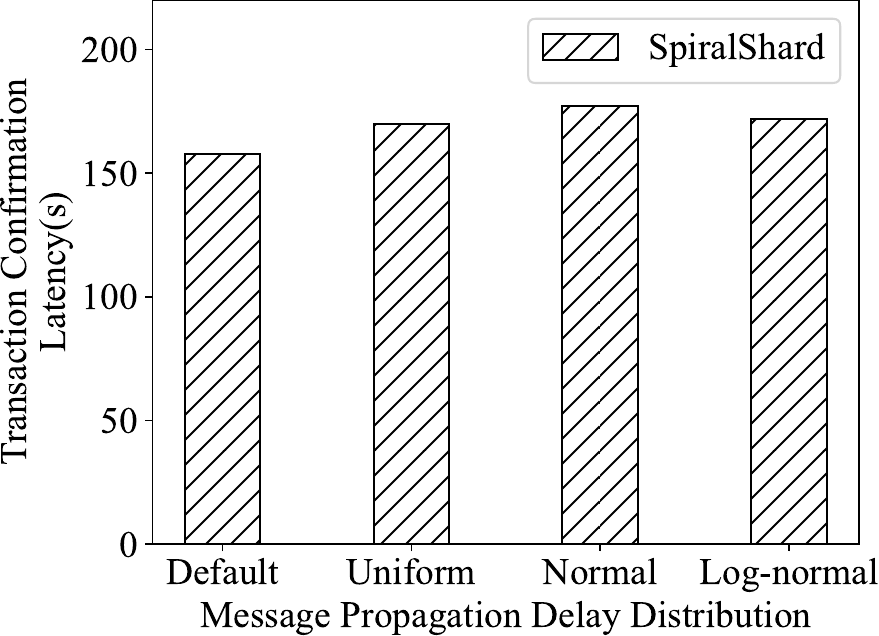}\label{subfig:latencyWithDelay}}
\centering
\vspace{-6pt}
\caption{Performance with different message propagation delay.}
\end{figure}

As shown in Figure~\ref{subfig:throughputWithDelay} and Figure~\ref{subfig:latencyWithDelay}, the additional delay results in a reduction in system throughput and an increase in system latency compared to the default network environment. 
This is primarily because the extra delay extends the time required to reach consensus. Nevertheless, the impact on system performance remains within acceptable bounds, with throughput loss and latency increase constrained to within 10\% and 12\%, respectively.
{\color{red}
These results also show that SpiralShard’s safety and liveness are not affected.
Under network delay attacks, the protocol continues to finalize blocks (albeit at a slightly reduced rate), ensuring the system does not stall and all valid transactions eventually confirm
}

%% file: Files/conclusion.tex
 \section{Conclusion}
In this paper, we propose SpiralShard, a highly concurrent and secure blockchain sharding system.
At its core, SpiralShard allows some shards to be corrupted.
To preserve the system's security and efficiency, we propose a Linked Cross-shard Endorsement protocol to eliminate forks and finalize non-conflicting blocks.
Benefiting from \ac{LCE}, SpiralShard can eventually allow some shards to tolerate less than $2/3$ fraction of malicious nodes ($1/3$ Byzantine nodes plus $1/3$ a-b-c nodes).
As a result, SpiralShard can configure smaller and more shards, enhancing system concurrency accordingly.
Finally, we implement SpiralShard based on Harmony and deploy a large-scale network for experiments to verify the superiority of SpiralShard.
The experimental results show that SpiralShard achieves up to 19$\times$ TPS compared with the Harmony baseline protocol.